  \documentclass[journal]{IEEEtran}

\usepackage{cite}
\usepackage{multicol,lipsum}
\usepackage{amsmath,amssymb,amsfonts}
\usepackage{algorithmic}
\usepackage{graphicx}
\usepackage{textcomp}
\usepackage{xcolor}
\usepackage{booktabs}
 \usepackage{hyperref}
\usepackage{enumerate}
\usepackage{algorithm}
\usepackage{multirow}
\usepackage{standalone}
\usepackage{tikz}
\usetikzlibrary{positioning}
\usetikzlibrary{decorations.markings}
\usepackage{caption}
\usepackage{subcaption}
 \usepackage{makecell}

\ifCLASSINFOpdf
\else
\fi
\begin{document}
%\markboth{  Dec.~2023}%

\title{ Advanced  Nonlinear    SCMA Codebook Design Based on Lattice Constellation}
 
 \author{ Qu Luo, \textit{Member, IEEE},  Jing Zhu, \textit{Member, IEEE},   Gaojie Chen, \textit{Senior Member, IEEE},   Pei Xiao, \textit{Senior Member, IEEE},   Rahim Tafazolli, and Fan Wang.

\thanks{
This work  was supported  by the U.K.    Engineering and Physical Sciences Research Council under Grant EP/P03456X/1 and  EP/X013162/1. (Corresponding author: Qu Luo)

Q. Luo,  J. Zhu,  G. Chen  and P. Xiao  are with 5G and 6G Innovation centre, Institute for Communication Systems (ICS) of University of Surrey, Guildford, GU2 7XH, U.K.  (e-mail: \{q.u.luo,  j.zhu,  gaojie.chen, p.xiao, r.tafazolli \}@surrey.ac.uk).
Fan Wang was   with Huawei Technologies   Company Ltd. 
 (email: fan.wang@huawei.com)
}} 
\maketitle

\begin{abstract}
%Sparse code multiple access (SCMA) on shared resources has been identified as a promising technology in machine type communication networks  to improve spectrum  efficiency and support massive connectivity.  
The design of efficient sparse codebooks in  sparse code multiple access (SCMA) system   have attracted tremendous research attention in the past few years. This paper proposes a novel nonlinear SCMA (NL-SCMA) that can subsume  the conventional SCMA system which is  referred to as linear SCMA,  as special cases  for downlink channels.   This innovative approach allows  a  direct mapping of users' messages to a superimposed codeword for transmission, eliminating the need of a codebook for each user. This mapping is referred to as nonlinear mapping (codebook) in this paper. 
 Hence, the primary objective is to  design the nonlinear mapping, rather than the linear codebook for each user.  We leverage the Lattice constellation to design the superimposed constellation due to its advantages such as the  minimum Euclidean distance (MED),  constellation  volume,  design flexibility and  shape gain. Then, by  analyzing the error patterns   of the Lattice-designed superimposed codewords  with the aid of the pair-wise error probability,  it is found that the MED of the proposed nonlinear codebook  is lower bounded by the ``single error pattern''. To this end, an error pattern-inspired codebook design is proposed, which can achieve    large MEDs of the nonlinear codebooks.  Numerical results show   that the proposed codebooks can  achieve lower error rate  performance over both Gaussian and Rayleigh fading channels than the-state-of-the-art linear codebooks.

%In the proposed nonlinear SCMA, the input message of all SCMA users are jointly mapped to a superimposed codeword for transmission, instead of individually  mapping each user's bits  to a codeword based on their predefined codebooks. 

\end{abstract}


% \fi
\IEEEpeerreviewmaketitle

\section{Introduction}
 
\IEEEPARstart{T}{he}  ever-increasing demand for higher data rates, improved spectral efficiency, and massive connectivity has driven the rapid evolution of wireless communication systems \cite{NGMA}. To meet these requirements, non-orthogonal multiple access (NOMA) technique has been envisioned as a promising technique  for  future wireless communication networks \cite{NGMA,luo2023design}. Distinguished from conventional orthogonal multiple access (OMA) technologies such as  time division multiple access and orthogonal frequency division multiple access, NOMA enables the transmission of multiple users over the same time-frequency resources \cite{luo2023design}.  
 Until now, many NOMA schemes have been proposed, including power-domain NOMA (PD-NOMA) \cite{PDNOMA1}, multiuser shared access (MUSA) \cite{MUSA}, pattern division multiple access (PDMA) \cite{ChenPDMA}, interleave-grid multiple access (IGMA) \cite{IGMA1}
and code-domain NOMA (CD-NOMA) \cite{ liu2021sparse}.  In particular, sparse
code multiple access (SCMA)  is a representative   CD-NOMA scheme, which carries out  multiplexing by employing carefully designed
codebooks/sequence \cite{LuoError}.   In SCMA, each user  is assigned an unique codebook, and the incoming message bits from the  user  are directly mapped to a multi-dimensional codeword, which   is intentionally  sparse so as to reduce the  decoding complexity by employing the  message passing algorithm (MPA) \cite{ChengEfficient,ChaiLDPC,ChenMIMOSCMA,YuanSCMA}.    The multi-dimensional codeword also provides   a constellation shape gain and consequently leads to better spectral efficiency when compared to other  CD-NOMAschemes such as  low density spreading
code division multiple access (LDS-CDMA) and low density
signature-orthogonal frequency division multiplexing (LDS-OFDM) \cite{ Hoshyar}.

\subsection{Related works}
A  fundamental research problem in SCMA is how to design efficient sparse codebook for low error rate transmission and by now, the SCMA codebook design has acquired the status of an open problem \cite{luo2023design,Huang,CaiMD,chen2020design,mheich2018design, yan2016top,XiaoCapacity,VikasComprehensive,Zhang,li2020design,wen2022designing}.   Current codebook design approaches typically follow a top-down methodology   by first constructing a   mother constellation (MC), upon which certain  user-specific  operators, such as phase rotation and permutation,   are applied to the MC to obtain codebooks for multiple users \cite{luo2023design,CaiMD,chen2020design,mheich2018design, yan2016top,XiaoCapacity,VikasComprehensive,Zhang}. 
The design key performance indicators (KPIs) for the MC and user-specific  operators are generally the the minimum product distance  (MPD) and the the minimum Euclidean distance  (MED) of the MC, and the MED of the superimposed constellation \cite{luo2023design}.   
In general, a large MED
leads to reliable detection in the Gaussian channel, whereas
a large MPD is preferred for robust transmissions in the
Rayleigh fading channels.

Following the top-down idea,   the pulse amplitude modulation (PAM)   constellation and golden angle modulation constellation (GAM) were considered   as the MC in \cite{CaiMD} and \cite{mheich2018design}, respectively,  where the MED of   the MC  is considered as  the  KPI  in downlink SCMA systems.  The $4$-PAM constellation was also adopted in \cite{XiaoCapacity,yan2016top} for large capacity, energy efficiency and mutual information gain.  It is noted that the same $4$-order  PAM  constellation was    employed as the basic constellation in \cite{XiaoCapacity,yan2016top}, thus their  resultant codebooks  exhibit    certain similarity.  In \cite{li2020design},    power-imbalanced codebooks were proposed by maximizing the MED of the superimposed constellation while maintaining  the  MPD of the MC larger than a threshold. 
In \cite{yu2015optimized}, the authors considered Star-QAM as the MC for enlarged   MED of   the superimposed codewords in downlink SCMA systems.  In \cite{chen2020design}, near-optimal  codebooks for different channel conditions  were investigated by choosing suitable MCs with large MPD or MED.    More recently,  a comprehensive approach to achieve near-optimal SCMA codebooks designs for $150\%$ and $200\%$ overloading factors under various channel environments was   studied in \cite{VikasComprehensive}. 

 There are yet other works combining constellation to design
SCMA codebooks.   The author in \cite{Zhang} proposed an uniquely decomposable constellation group with amplitude and phase-shift keying modulation. In addition,  a joint design of  codebook and peak-to-average power ratio (PAPR) reduction based on the partial transmit sequence technique was proposed in \cite{ChenPAPRSCMA}.  Different from the above works, which mainly design  the codebooks over Gaussian or Rayleigh fading channels, the recent work in \cite{luo2023design} studied the codebook design in Ricain fading channels, where a novel class of low-projection SCMA codebooks for ultra-low decoding complexity  was  developed. As revealed by many SCMA works \cite{luo2023design,Huang,chen2020design, yan2016top,XiaoCapacity,VikasSCMA,VikasComprehensive,Zhang},  the distance properties of the superimposed constellation, such as the MED of the superimposed constellation, the MED at each resource node, the volume of the constellation at each subcarrier, the PAPR of the  superimposed constellation,   play an essential role in improving the  performance of SCMA systems.

\subsection{Motivation and Contribution}

 One fundamental characteristic of the above SCMA codebook design is the utilization of an  MC  to generate the multi-dimensional codebook based on the top-down principle.  Thus, the performance of the codebook relies heavily on the selection of an MC and the user-specific constellation operators  \cite{luo2023design,Huang,chen2020design, yan2016top,XiaoCapacity,VikasSCMA,VikasComprehensive,Zhang}. 
Numerous studies on SCMA have highlighted the crucial role of the  the superimposed constellation, such as the MED  \cite{Huang,chen2020design, yan2016top,VikasComprehensive,Zhang}, the MED at each subcarrier  \cite{chen2020design,VikasComprehensive}, the volume of the constellation at each subcarrier  \cite{yan2016top,VikasSCMA}. However, designing a codebook that simultaneous achieves these objectives poses a considerable challenge to the existing design schemes. An alternative and more ambitious approach is to directly design the superimposed constellation with desirable characteristics instead of following a top-down based principle.  This is reasonable as the user  bit message can be uniquely determined by the superimposed constellation as long as    the superimposed constellation is well designed \cite{luo2023design}.

In this paper,  a novel nonlinear SCMA (NL-SCMA)   is proposed, where the input message of SCMA users are directly mapped to the superimposed codeword according to a specific nonlinear mapping rule. The mapping process is referred to as nonlinear codebook (NLCB) in this paper, whereas the conventional SCMA codebook is thus called linear codebook (LCB). Motivated by the superiority of the Lattice constellation, such as the large shape gain, large MED, good design flexibility and superior performance \cite{ErezLattices}, we employ the Lattice constellation to design the superimposed constellation. As such, the resultant superimposed constellation   can also enjoy these benefits \cite{ErezLattices}. The main contributions of this work can be summarized as follows:

 \begin{itemize}

\item We propose a NL-SCMA system that directly maps the input message of SCMA users to the superimposed codeword. The proposed NL-SCMA can serve as a general framework and
subsumes the existing   linear SCMA systems as special cases.

\item A  constellation partition  scheme is proposed to design the  superimposed constellation at each subcarrier based on the Lattice constellation, where the partitioned constellation is called Lattice code. 

\item We propose an error-pattern inspired  method to design the nonlinear mapping  between the input message and the  multi-dimensional superimposed codeword by exploiting  the unique characteristic of the pair-wise error probability (PEP)   of    the proposed NL-SCMA. Hence, the proposed scheme is referred to as error-pattern inspired NLCB design.

% \item A linear codebook is also proposed in this paper, whose superimposed constellation at each subcarrier is a Lattice code and thus the proposed linear codebook  can enjoy the benefits of the  Lattice constellation. 

\item We conduct extensive numerical experiments to show the superiority of the proposed linear and NL-SCMA codebooks. Numerical results indicate  that the proposed codebooks can simultaneous achieve lower error rate performance in uncoded and coded systems and  large MED  than the-state-of-the-art codebooks. 

 \end{itemize}

\subsection{Organization}
 The remainder of this paper are outlined as follows.
Section II introduces the system model of linear SCMA and the proposed NL-SCMA systems. The SCMA codebook design  KPIs
are illustrated in Section III,  followed by the brief introduction of existing linear SCMA designs. In Section IV, the detailed design of the NL-SCMA   and LCBs are discussed. Section V is the numerical results, followed by the conclusions in Section VI

\subsection{Notation}

  $\mathbb{C}^{k\times n}$ and $\mathbb{B}^{k\times n}$ denote the $(k\times n)$-dimensional complex and binary matrix spaces, respectively. ${{\mathbf{I}}_{n}}$ denotes an $n \times n $-dimensional  identity matrix.   $\text{diag}(\mathbf{x})$ gives a diagonal matrix with the diagonal vector of $\mathbf{x}$. $(\cdot)^\mathcal T$, $(\cdot)^ \dag $ and $(\cdot)^\mathcal H$ denote the transpose, the conjugate and the Hermitian transpose operation, respectively.  $\|\mathbf{x}\|_2$ and $|x|$ return the Euclidean norm of vector $\mathbf{x}$ and the absolute value of $x$, respectively. $\mathcal{CN}(0,1) $ denotes the complex distribution with zero mean and unit variance.   
 
\section{System Model}

\subsection{SCMA Communication Model}
\begin{figure}
     \centering
  \includegraphics[width=0.38 \textwidth]{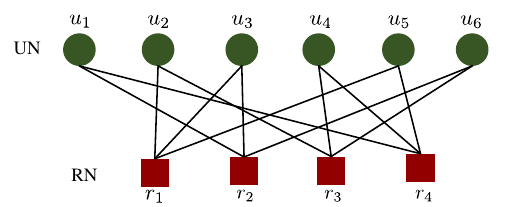}
         \caption{Factor graph representation of a SCMA system.}
         \label{facotGraph}
\end{figure}

We consider a downlink SCMA system, where $J$ users (UEs) communicate over  $K$ orthogonal subcarriers. The overloading factor defined as $J/K$ is   larger than one. In  SCMA, each user is assigned with  an unique codebook, denoted by  $ \boldsymbol {\mathcal {X}} _{j}=\left\{\mathbf{x}_{j, 1}, \mathbf{x}_{j,2}, \ldots, \mathbf{x}_{j,M}\right\} \in \mathbb {C}^{K \times M}, j \in\{1,2, \ldots, J\}$, where $M$ is the codebook size and $\mathbf{x}_{j,m}$ is the $m$th codeword with   a dimension of $K$.  During transmissions,  each user   maps $\log_2\left(M\right)$   binary bits to a length-$K$ codeword $ \mathbf {x} _{j}$ drawn from the  $ \boldsymbol {\mathcal {X}}_{j}  $.  The mapping process can be expressed as \cite{MultitaskSCMA}
\begin{equation}
\small
\label{b2x}
    f_{j}: \mathbb{B}^{\log_{2}M \times 1} \rightarrow \boldsymbol{\mathcal { X }}_{j} \in \mathbb{C}^{K \times M}, \text { i.e., } \mathbf{x}_{j}=f_{j}\left(\mathbf{b}_{j}\right),
\end{equation}
where $\mathbf{b}_{j}=\left[b_{j, 1}, b_{j, 2}, \ldots, b_{j, \log _{2} M}\right]^{\mathcal {T}} \in \mathbb{B}^{\log _{2} M \times 1}$ represents the binary message input for the $j$th user. The $K$-dimensional complex codewords in the SCMA codebook are sparse vectors with $N$ non-zero elements and $N< K$.   The factor graph  can be used to represent the connections between user nodes (UNs) and resource nodes (RNs) in SCMA. The sparse structure of the $J$ SCMA codebooks can also be represented by the indicator  matrix   $\mathbf {F}_{K \times J}   \in \mathbb {B}^{K\times J}$.  An element of  ${\bf{F}}$ is defined as ${f_{k,j}}$ which takes the value of $1$ if and only if  UN  $u_j$ is connected to   RN  $r_k$ and 0 otherwise.  In this paper, the following   indicator matrix  with $J=6$, $K=4$ and $N=2$ is considered \cite{MultitaskSCMA}: 
  \begin{equation} 
 \label{Factor_46}
 \small
 {{\mathbf{F}}_{4\times 6}}=\left[ \begin{matrix}
   0  & 1 & 1 & 0 & 1 & 0  \\
   1 & 0 &1& 0 & 0 & 1\\
   0 & 1& 0 & 1 & 0 & 1 \\
   1 & 0 & 0 &1 & 1 & 0  \\
\end{matrix} \right],
  \end{equation}
where the  factor graph representation is shown in Fig. \ref{facotGraph}. 

 Let $\mathbf {c}_{j}$ be a length-$N$ vector drawn from  $  {\mathcal C}_{j}\subset \mathbb {C}^{N \times 1}$, where $ {\mathcal C}_{j}$ is obtained  by removing all the zero elements  in  $ { \mathcal X}_{j}$.  We further define the mapping from $\mathbb{B}^{\log_2M\times 1} $ to  $ {\mathcal C}_{j}$ as \cite{mheich2018design}
\begin{equation} \label{SCMAmapping}
\small
q_{j}:\mathbb {B}^{\log _{2}M\times 1}\mapsto   \boldsymbol {{\mathcal C}}_{j}, \quad {~\text {i.e., }}\mathbf {c}_{j}=q_{j}(\mathbf {b}_{j}).
\end{equation}
Thus, the corresponding  SCMA mapping  $v_j$ can be expressed as
\begin{equation}
\label{scmaMapping}
\small
v_{j}:    \mathbf {V}_{j}q_{j}  \mapsto \boldsymbol{\mathcal { X }}_{j}, \quad {~\text {i.e., }}\mathbf {x}_{j}=\mathbf {V}_{j}q_{j}(\mathbf {b}_{j}),
\end{equation}
where $\mathbf {V}_{j} \in \mathbb {B}^{K \times N} $ is a  mapping   matrix that maps the $N$-dimensional vector  to a $K$-dimensional sparse SCMA codeword. Based on the factor graph, $\mathbf {V}_{j}$ 
can be constructed  according to the position of the ‘$0$’ elements of  ${{\mathbf{f}}_{j}}$ by inserting  all-zero row vectors into the identity matrix ${{\mathbf{I}}_{N}}$. For example, we have
 ${{\mathbf{V}}_{1}}=\left[ \begin{matrix}
  0 & 1  & 0 & 0  \\
  0 & 0 &  0 & 1  \\
\end{matrix} \right]^{\mathcal T} \text{and} \quad {{\mathbf{V}}_{2}}=\left[ \begin{matrix}
  1 & 0  & 0 & 0  \\
  0 & 0 &  1 & 0  \\
\end{matrix} \right]^{\mathcal T},$ and  $\mathbf {V}_{j}, j= 3,4,5$ and $6$ can be generated in a similar way.

For a downlink SCMA system, the transmitted codeword  of each user is first superimposed at the base station, leading to a superimposed codeword, denoted by $\mathbf {w} =[w_{1},w_{2}, {\dots },w_{K}]^\top \in \mathbb {C}^{K\times 1}$. Namely, we have \cite{mheich2018design}
\begin{equation} 
\small
\label{x2w}
\mathbf {w} = \sum \limits ^{J}_{j=1}\mathbf {x}_{j}.
\end{equation}

\color{black}

\subsection{The Proposed  NL-SCMA}

 \begin{figure}
     \centering
  \includegraphics[width=0.45 \textwidth]{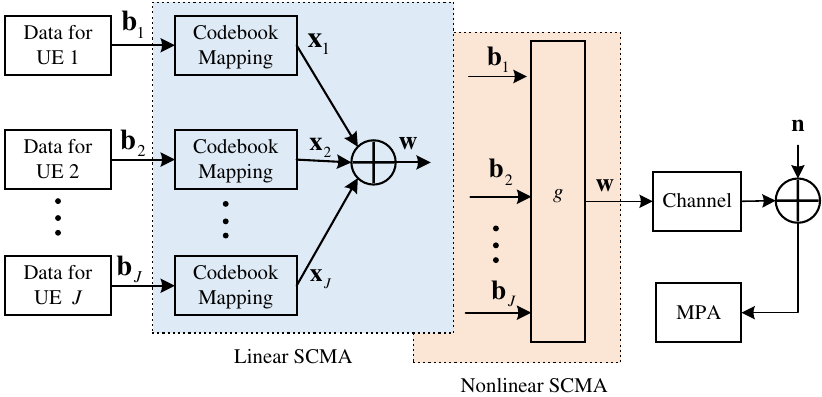}
         \caption{ Block diagram for the downlink linear and nonliner SCMA systems.}
         \label{Figsys}
\end{figure}

In existing SCMA systems, the transmitted vector $\mathbf{w}$ is obtained by combining the codewords of $J$  users,  which will constitute a superimposed constellation, denoted by $\boldsymbol{\Phi}_{M^J} \in \mathbb{C}^{K \times M^J}$.  Based on (\ref{x2w}),  the superimposed constellation can be obtained as 
\begin{equation} 
\small
\label{super}
\mathbf {\Phi}_{M^J} =\left \{ \sum \limits ^{J}_{j=1}\mathbf {x}_{j} \Big\vert \forall \mathbf {x}_{j} \in \boldsymbol{\mathcal { X }}_{j}, \forall j\right \}.
\end{equation}
As a matter of fact, the mappings described by   (\ref{scmaMapping}) and (\ref{x2w}) are not necessary in downlink SCMA systems. Alternatively, the input message can be directly mapped to a vector $\mathbf{w}$. This leads to the definitions of the linear SCMA and NL-SCMA as follows:

  \textit{Definition 1: If the transmitted codeword   $\mathbf w $ is obtained through  (\ref{b2x}) and (\ref{x2w}), then it is called a linear SCMA system and the codebook $\boldsymbol{\mathcal { X }} = \{\boldsymbol{\mathcal { X }}_{1},\boldsymbol{\mathcal { X }}_{2},\ldots,\boldsymbol{\mathcal { X }}_{J} \}$ is said to be a LCB set.  On the other hand,  if the transmitted codeword   $\mathbf w $ is obtained by the following non-linear mapping rule:}
\begin{equation}
\small
\label{b2w}
    g: \mathbb{B}^{\log _{2} M \times J} \rightarrow \mathbf{  w} \in \mathbb{C}^{K \times 1}, \text { i.e., } \mathbf{w} = g\left(\mathbf{B}\right),
\end{equation}
where $ \mathbf{B} = [ \mathbf b_1,\mathbf b_2, \ldots, \mathbf b_J  ]$, then it is referred as a NL-SCMA system  and  the mapping  $g$ is said to be a NLCB.

Fig. \ref{Figsys} shows an example of the linear SCMA and the proposed NL-SCMA systems.   It is noted that the proposed NL-SCMA can  serve as a general framework and subsume  the existing linear SCMA systems as special cases.
 Denote $\mathbf {h}_l  \in \mathbb {C}^{K\times 1}$ as the channel vector from the base station to the $l$th UE. The received signal at the $l$th UE  can be expressed as 
\begin{equation} 
\small
\label{signal_model}
 \mathbf {y}_{l}  = \left\{\begin{matrix}
   \mathrm {diag}(\mathbf {h}_{l}) \sum \limits ^{J}_{j=1}\mathbf {x}_{j} +\mathbf {n}_{l} & \text{Linear SCMA, } \\ 
  \mathrm {diag}(\mathbf {h}_{l}) g\left(\mathbf{B}\right) +\mathbf {n}_{l} & \text{Nonlinear   SCMA, } 
\end{matrix}\right.
\end{equation}
where $\mathbf {n}_{l}$ denotes the    noise vector at the $l$th UE that has $\mathcal{CN}(0,N_0)$   entries. For simplicity, the  subscript $l$ in (\ref{signal_model}) is omitted whenever no ambiguity arises.

 It is noted that the proposed NL-SCMA does not increase computational complexity compared to conventional SCMA systems. It requires the base station to store the superimposed constellation, which has $K \times M^J$ complex-valued elements. In contrast, linear SCMA systems require the base station to store the codebooks of 
$J$ users, which have 
 $JMK$  complex-valued elements.

\subsection{  NL-SCMA Detection}
 Given the received signal   $\mathbf y$ and $\mathbf h$, the optimum maximum a posterior probability   detector finds  the   $\widetilde {\mathbf{B}}$ that maximizes the a posterior probability of the transmitted ${\mathbf{B}}$, i.e., ${\text {Pr}}(\boldsymbol {\mathbf{B}}|\boldsymbol {y})$, by \cite{weijieFTN}
\begin{equation}
\small
\label{map1}
     \widetilde {\mathbf {B}}=\arg \underset{{\mathbf {B}}\in \mathbb {B}^{\log _{2} M \times J}}{\max } {\text {Pr}}({\mathbf {B}}|\boldsymbol {y}, g).
\end{equation}

According to Bayes’ rule, we have
\begin{equation}
\small
\label{map2}
\begin{aligned}
{\text {P}}(\mathbf {B}|\mathbf {y},g) &=\frac{{\text {Pr}}(\mathbf {y}|\mathbf {B},g){\text {Pr}}(\mathbf {B})}{{\text {Pr}}(\mathbf {y})} \propto {\text {Pr}}(\mathbf {y}|\mathbf {B},g){\text {Pr}}(\mathbf {B}),
\end{aligned}
\end{equation}
where ${\text {P}}(\mathbf {B})=\prod _{j=1}^{J}{\text {Pr}}(\mathbf b_j)$ is the joint prior  probability mass function.   
Under the Gaussian assumption of $\mathbf n$,  we have
\begin{equation} 
\small
\label{posterP}
\begin{aligned}
{\text {P}}(\mathbf {y}|\mathbf{B},g) = \frac{1}{\sqrt{2\pi N_0}} \exp \left(-\frac{1}{2N_0}\Vert \mathbf {y}-  \mathrm {diag}(\mathbf {h})g( \mathbf{B})\Vert ^2\right). 
\end{aligned}
\end{equation}

Solving problems (\ref{map1})-(\ref{posterP})   has exponential complexity. Thanks to the sparsity of the codewords, the solution of this problem can be approximated by an iterative decoding algorithm. 
Following the factor graph and the principles of the sum-product
algorithm, the information transmitted from the $k$th RN  ($r_k$) to the $j$th UN  ($u_j$) 
  can be expressed as \cite{LiuYushaSCMA}
\begin{equation}
\small
\delta _{r_k\rightarrow u_j}^{t} (\mathbf b_j)= \sum _{  \widetilde{\mathbf{b } }_j \in \mathcal B}  {\text {P}}(\mathbf {y}| \widetilde{\mathbf{B}},g) 
 \prod _{l \in \xi_k \backslash \{j \}} \eta _{u_l\rightarrow r_k}^{t-1} , \\
\end{equation}
 where $ \widetilde {\mathbf{B}} = [ \widetilde {\mathbf b}_1,\widetilde {\mathbf b}_2, \ldots, \widetilde {\mathbf b}_J  ] \in \mathbb B^{\log_2M \times J}$,   $ \mathcal B $  denotes   all possible $\widetilde {\mathbf{B}}$  matrices in which $\widetilde {\mathbf b}_j \neq    \mathbf b_j$,   and $\xi_k $ denotes the set of user indices which utilizes the $k$th
resource.   By contrast, the information conveyed from UN $u_j$  to RN $r_k $  is the product of the information collected from the UNs connected to RN $r_k$   excluding UN $u_j$, which can be expressed as
\begin{equation}
\small
\eta _{u_j\rightarrow r_k}^{t}=  \prod _{m  \in \zeta_j \backslash k} \delta _{r_m\rightarrow u_j}^{t-1},
\end{equation}
where $\zeta_j $  is the set of nonzero’s position in the $j$th column of  ${{\mathbf{F}}_{K\times J}}$.

\section{Sparse Codebook Design: Design Metric}

In this section, we  present the  KPIs that are considered for designing advanced codebooks.    To better understand the proposed codebook design principle, we  recall briefly the state-of-the-art codebook design scheme.

\subsection{SCMA Codebook Design KPIs}

   We start by introducing the MED and MPD in a linear SCMA system.   The  MED and MPD  of an $N$-dimensional constellation  $\mathbf{ A} = [\mathbf a_1, \mathbf a_2, \ldots, \mathbf a_M] \in \mathbb C^{N \times M}$   are defined as \cite{li2020design,mheich2018design} 
\begin{equation} 
\label{dmin_mc}
\small
\text{MED} \left (\mathbf{A} \right) \triangleq  \underset {\substack {1 \leq p,  q \leq M, \\ p \neq q}}{\min }    \Arrowvert \mathbf {a}_{p} -\mathbf {a}_{q} \Arrowvert, 
\end{equation}
and
\begin{equation} 
\small
\text{MPD} \left (\mathbf{A} \right) \triangleq \underset {p\neq q}{\min } \prod _{n \in \rho(\mathbf {a}_{p}, {\mathbf {a}_{q}})} |a_{n,p} -a_{n,q}|^{2},
\end{equation}
respectively, where  $ {a}_{n,p}  $ denotes the $n$th entry  of  $\mathbf {a}_{p} $ and $\rho (\mathbf {a}_p, {\mathbf {a}}_q)$ denotes the set of  indices in which $ {a}_{n,p} \neq {{a}}_{n,q}$.  The MED of  $ \boldsymbol{\Phi}_{M^J}$ is obtain by calculating  ${{{M}^{J}}\left( {{M}^{J}}-1 \right)}$ mutual distances between ${{M}^{J}}$ superimposed codewords \footnote{ The metric of the MED and MPD of the MC, and the MED of the superimposed constellation are derived based on the PEP analysis of SCMA. We refer the readers  to \cite{luo2023design} and \cite{chen2020design} for more details about the PEP formulation. }, i.e.,
\begin{equation} 
\label{dmin_mc}
\small
\text{MED} \left ( \boldsymbol{\Phi}_{M^J} \right)  \triangleq  \underset {\substack {1 \leq p,  q \leq M^J, \\ p \neq q}}{\min }    \Arrowvert \mathbf {w}_{p} -\mathbf {w}_{q} \Arrowvert.
\end{equation}

 Analogous to the MED in linear SCMA systems, the MED of the superimposed codewords in the proposed NL-SCMA can be represented as
\begin{equation} 
\label{dmin_mc}
\small
\text{MED} \left ( \boldsymbol{\Phi}_{M^J} \right)  \triangleq  \underset {\substack {  \forall \mathbf B, \forall \widetilde{\mathbf B}, \\ g(\mathbf B) \neq  g(\widetilde{\mathbf B}) }}{\min }    \Arrowvert  g(\mathbf B) -g(\widetilde{\mathbf B}) \Arrowvert.
\end{equation}

 The MPD of the NL-SCMA will be introduced along with the NL-CB design in Section IV.   In general, a large $\text{MED} \left ( \boldsymbol{\Phi}_{M^J} \right)$   leads to reliable detection in the Gaussian channels, whereas a large
 $\text{MPD} \left (\boldsymbol {\mathcal{X}}_j \right)$ is preferred for robust transmissions in the Rayleigh fading channels \cite{luo2023design}.

 \begin{figure}
     \centering
     \begin{subfigure}[b]{0.24\textwidth}
         \centering
  \includegraphics[width=1 \textwidth]{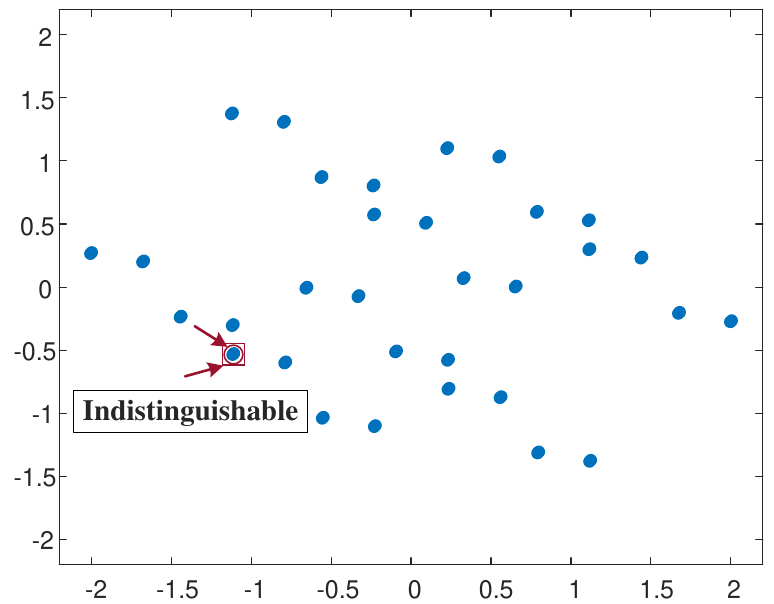}
         \caption{ GA  codebook   $ \text{MED} \left ( \boldsymbol{\Phi}_{M^{{d_f}}}(4) \right) $} \cite{GAKlimentyev}. 
         \label{LattHex}
     \end{subfigure}
     \hfill
     \begin{subfigure}[b]{0.24\textwidth}
         \centering
  \includegraphics[width= 1 \textwidth]{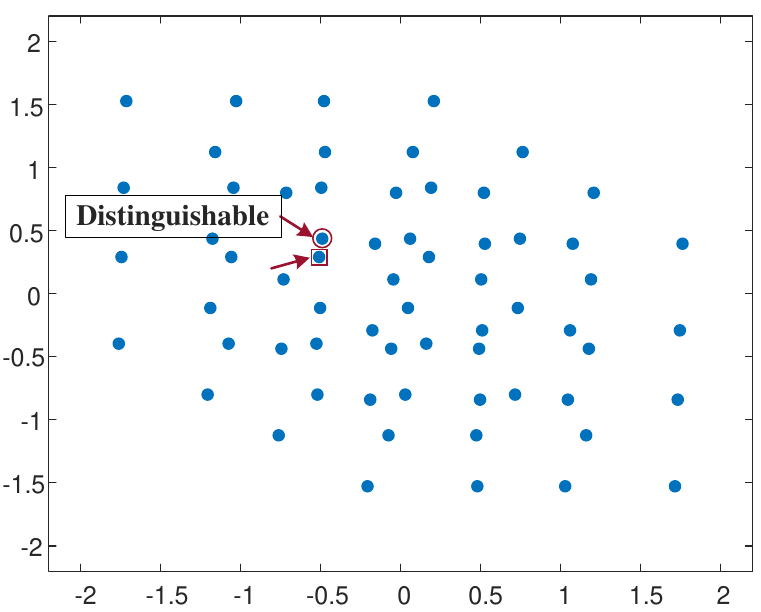}
         \caption{   Huawei codebook   $ \text{MED} \left ( \boldsymbol{\Phi}_{M^{{d_f}}}(3) \right) $}  \cite{huawei}.
         \label{LattQAM}
     \end{subfigure}
              \caption{Examples of the superimposed constellations.}
              \label{LatticeCon}
\end{figure}

\subsection{The Top-Down Based Multi-Stage Codebook Design }

It is worth mentioning that current codebook design approaches typically follow a top-down methodology   by first constructing an  MC, upon which certain  phase rotations  are applied to the MC to obtain codebooks for multiple users \cite{CaiMD,chen2020design,mheich2018design,yan2016top,XiaoCapacity,VikasSCMA,VikasComprehensive,Zhang}.   These phase rotations aim to enhance the distance metric between superimposed codewords, thereby improving overall codebook performance.  For $K=4, J=6, N=2$, a Lattice based phase rotation   matrix is widely considered, which takes the following expression:
  \begin{equation} 
 \label{signature_46}
 \small
 {{\mathbf{\Theta}}_{4\times 6}}=\left[ \begin{matrix}
   0  & e^{j\theta_3} & e^{j \theta_1} & 0 &  e^{j\theta_2} & 0  \\
   e^{j\theta_2} & 0 & e^{j\theta_3} & 0 & 0 & e^{j\theta_1}\\
   0 & e^{j\theta_2}& 0 & e^{j\theta_1} & 0 & e^{j\theta_3} \\
   e^{j\theta_1} & 0 & 0 & e^{j\theta_2} & e^{j\theta_3} & 0  \\
\end{matrix} \right],
  \end{equation}
where $\theta_i \in [-\pi,\pi]$ is the  phase  rotation. The steps  involved in the codebook design for AWGN or Rayleigh fading channels are summarized below.

 \textbf{Step 1: } Select an one-dimensional basic constellation, denoted as $\mathbf a_{0} \in \mathbb C^{ M \times 1}$. 
 
 \textbf{Step 2: }   The MC, defined as  $\mathbf{A}_{ \text{MC}}$,  is obtained by the   repetition and permutation  of  $\mathbf a _0$.   Let   $\boldsymbol {\pi }_{n}  $ denote  the permutation mapping of the $n$th dimension,   then the $N$-dimensional MC  can be obtained as 
\begin{equation}
 \small
 \label{permu}
\mathbf{A}_{ \text{MC}} = \big[  { \boldsymbol {\pi }_{1} ({ \mathbf a _0}), \boldsymbol {\pi }_{2} ({ \mathbf a _0}), \ldots,  \boldsymbol {\pi }_{N}   ({ \mathbf a _0})} \big]^{\mathcal T}.
\end{equation}
The permutation criteria are     the $\text{MED} \left (\mathbf{A}_{ \text{MC}} \right) $ and $\text{MPD} \left (\mathbf{A}_{ \text{MC}} \right) $ for Gaussian and Rayleigh fading channels, respectively  \cite{luo2023design}.
 
 \textbf{Step 3: } The phase rotation angles are then applied to the $\mathbf{A}_{ \text{MC}}$ to design the codebook for different users.  Specifically,  the $j$th user's codebook is generated by  $\boldsymbol{\mathcal X}_{j} = \mathbf {V}_{j} \mathbf{\Theta}_j\mathbf{A}_{ \text{MC}}$, where  $\mathbf{\Theta}_j$   is the phase operation matrix of $j$th user that determined by (\ref{signature_46}). Specifically, $\mathbf {\Theta}_{j}$ 
can be constructed  by applying the diagonal operation  to the non-zero elements  of the $j$th column in (\ref{signature_46}).  For example,  we have
\begin{equation}
\small
{{\mathbf{\Theta}}_{1}}=\left[ \begin{matrix}
  e^{j\theta_2} & 0    \\
  0 & e^{j\theta_1}    \\
\end{matrix} \right]^{\mathcal T}, \quad {{\mathbf{\Theta}}_{2}}=\left[ \begin{matrix}
  e^{j\theta_3} & 0    \\
  0 & e^{j\theta_2}    \\
\end{matrix} \right]^{\mathcal T}.
\end{equation}
 
 Denote $\boldsymbol{\Phi}_{M^{d_f}}(k)$  as the superimposed constellation at the $k$th subcarrier, and let $d_f = \vert\xi_k\vert$, $E_{\max}(\boldsymbol{\Phi}_{M^{d_f}}(k))$.  The rotation angle should be optimized to improve distance profile of the superimposed codewords.   Specifically,  the optimum rotation angles $ \theta_i  =1, 2,\ldots, d_f$  are determined by maximizing the   $  \text{MED} \left ( \boldsymbol{\Phi}_{M^J} \right)  $,   $ \text{MED} \left ( \boldsymbol{\Phi}_{M^{{d_f}}}(k) \right), $  for Gaussian and Rayleigh fading channels, respectively \cite{luo2023design,chen2020design}.

 In downlink SCMA systems, the distance properties of the superimposed constellation, which are generally determined by the selection of an MC and rotation angles, have a significant impact on SCMA performance. On one hand, the top-down multi-stage codebook design is a sub-optimal approach, as it has been  pointed out that the optimal way is to directly design the superimposed constellation. On the other hand, the search space for rotation angles and parameters in MC designs is generally large. Therefore, it is quite challenging to obtain an MC and rotation angles that simultaneously yield large   $  \text{MED} \left ( \boldsymbol{\Phi}_{M^J} \right), $    
     $\text{MED} \left ( \boldsymbol{\Phi}_{M^{{d_f}}}(k) \right)$, and   $\text{MPD} \left (\mathbf{A}_{ \text{MC}} \right)$.
Fig. \ref{LatticeCon} shows an example of the  $\boldsymbol{\Phi}_{M^{d_f}}(4)$ and $\boldsymbol{\Phi}_{M^{d_f}}(3)$ of the GA codebook \cite{GAKlimentyev} and Huawei codebook  \cite{huawei}, respectively.  The GA codebook is generated by first selecting a mother MC, followed by the application of phase rotation and power scaling, which are optimized using a genetic algorithm, to generate the codebook for different users.   
  In Fig. \ref{LatticeCon}(a), it is apparent that multiple constellation points overlap at the same position, rendering them indistinguishable in the fourth dimension. While these codewords can still be distinguished by other dimensions, this results in a sacrifice of diversity gain in the fourth dimension \cite{chen2020design}.  In this regard, the codebook  $\boldsymbol{\Phi}_{M^{d_f}}(3)$ shown in Fig. \ref{LatticeCon}(b) is more preferable than  $\boldsymbol{\Phi}_{M^{d_f}}(4)$ depicted in Fig.  \ref{LatticeCon}(a) as all its elements are  separated. However, Fig.  \ref{LatticeCon}(b) still experiences a small   $\text{MED}\left(\boldsymbol{\Phi}_{M^{d_f}}(k)\right)$.  In this paper, a codebook is considered to be a ``full diversity" codebook if its superimposed constellation satisfies  $\text{MED}\left ( \boldsymbol{\Phi}_{M^{d_f}}(k) \right ) >0, 1 \leq k \leq K$. In addition to satisfying this condition, a large  $\text{MED}\left(\boldsymbol{\Phi}_{M^{d_f}}(k)\right)$ is also desirable for enhancing the diversity gain.

 The objective of this work is to design  an advanced class of   NLCBs that exhibit low error rates in both Gaussian and Rayleigh fading channels, while simultaneously achieving desirable characteristics such as  full diversity.
  
\section{The Proposed NLCB Design }

  Different from the ``top-down'' based design principles, we follows a  ``down-top'' based  idea to construct the   NL-SCMA codebooks based on the lattice constellation.

 \begin{figure}
     \centering
     \begin{subfigure}[b]{0.32\textwidth}
         \centering
  \includegraphics[width=1.1 \textwidth]{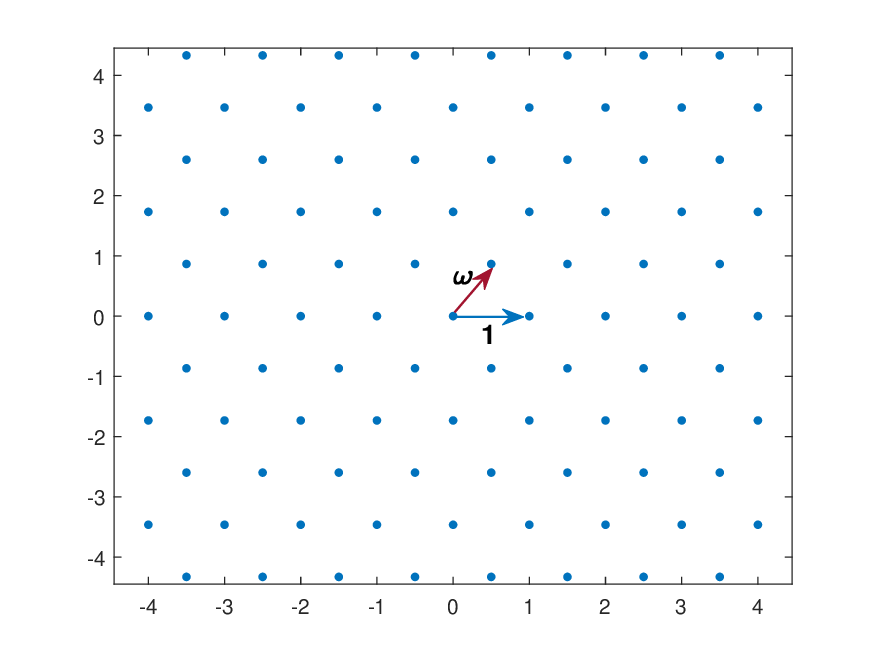}
         \caption{Lattice constellation of the  Eisenstein integers.}
         \label{LattHex}
     \end{subfigure}
     \hfill
     \begin{subfigure}[b]{0.32\textwidth}
         \centering
  \includegraphics[width= 1.1 \textwidth]{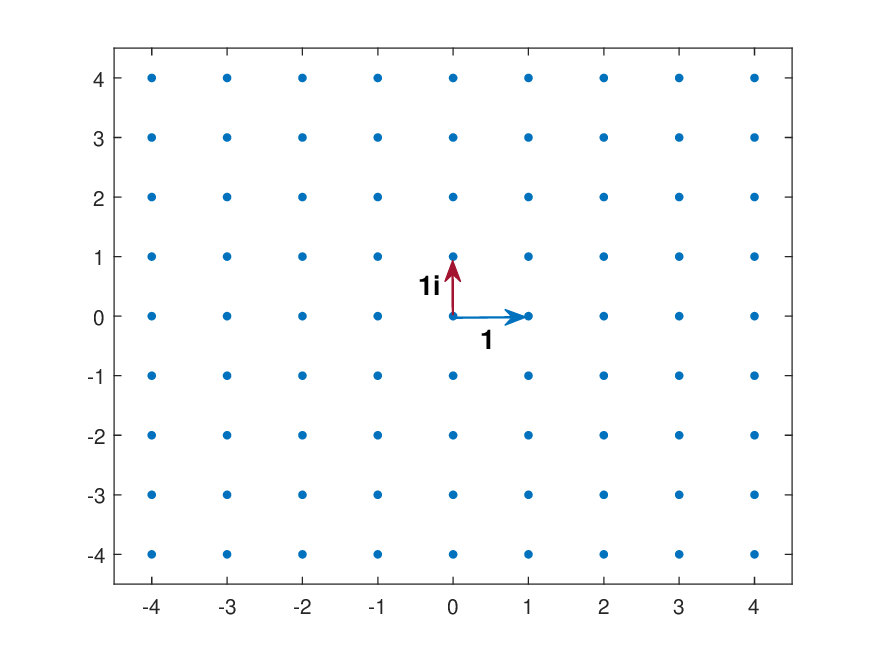}
         \caption{Lattice constellation of the Gaussian integers.}
         \label{LattQAM}
     \end{subfigure}
              \caption{The employed lattice constellations.}
              \label{LatticeCon}
\end{figure}

\subsection{The Proposed Non-Linear SCMA Codebook }

The first step in the proposed approach involves designing the superimposed constellation   with desirable distance properties. Then, the objective is to determine the non-linear mapping between users' input bits and the superimposed constellation.   The main steps involved in
the codebook design for both AWGN and Rayleigh fading channels
are   briefly summarized  follows:

\begin{itemize}
  \item   Designing an one-dimensional overlapped constellation $\boldsymbol{ \mathcal S} \in \mathbb C^{M^{d_f} \times 1}$ with full  diversity and  large  MED. Then, 
 $\boldsymbol{ \mathcal S} $      is utilized to     design   the superimposed constellation.
  \item  For each RN, design  the mapping between  users' input bits to the overlapped constellation  $\boldsymbol{ \mathcal S} $.
  \item   Optimizing the overall mapping to attain large $  \text{MED} \left ( \boldsymbol{\Phi}_{M^J} \right)$.
\end{itemize}

\subsubsection{Designing $\boldsymbol{ \mathcal S}$ based on lattice code} 
   A lattice constellation is a regular, repeating arrangement of points in a multidimensional space. It can be thought of as a grid that extends infinitely in all directions within that space \cite{ForneyMultidimensional}. It has the advantages of good minimum MED, constellation volume, design flexibility, and shape gain. Hence, in this paper, we consider lattice codes for the overlapped constellation design. The design process primarily consists of the following two steps. 

\textbf{Step 1a): Generate the lattice constellation}

To begin with, we introduce the following definition:

\textit{Definition 2: A lattice is defined as  the (infinite) set of points in an $n$-dimensional space given by all linear combinations with integer coefficients of a basis set of up to $n$ linearly independent vectors \cite{ForneyMultidimensional}.   Denote $ \mathbf   \Lambda \in \mathbb C^{n}$ as the $n$-dimensional complex-valued Lattice constellation,   then $ \mathbf   \Lambda$ can be defined in terms of a generator matrix $\mathbf G \in  \mathbb C^{n \times 
 2n} $, whose columns are the basis vectors:
\begin{equation}
\label{latt}
    \small
   \mathbf   \Lambda = \left\{ \lambda = \mathbf G \mathbf z \vert  \forall \mathbf z \in \mathbb Z^{2n} \right\}. 
\end{equation}}
A lattice code is then defined by the  finite  set of lattice points within a certain region.

In this paper, we consider two lattice constellations, namely the Hexagonal constellation   and the Gaussian lattice constellation \cite{ForneyMultidimensional}.  The Hexagonal constellation can be generated based on  the Eisenstein number, which takes the following expression  
\begin{equation}
    \small
    \nu = \exp\left(\frac{2\pi i}{3}\right),
\end{equation}
where $i=\sqrt{-1}$.  The above Eisenstein number is a root of $x^2+x+1 = 0$.  For an one-dimensional complex lattice constellation, $\mathbf G$ is a vector.  As a matter of fact,   by letting  $\mathbf  G = \left [ 1, \nu \right ]$, (\ref{latt})  forms the  one-dimensional hexagonal lattice in the complex domain \cite{LiuLDS}.    In the theory of data packing, a hexagonal lattice is viewed as the best lattice for its largest packing distance. 

 Gaussian lattice constellation is also known as the Gaussian integers or the quadrature amplitude modulation (QAM) constellation. The one dimensional complex Gaussian integers can be generated  by  letting $\mathbf G = \left [ 1, 1i \right ]$. Fig. \ref{LatticeCon} shows an example of the    of hexagonal   and  Gaussian lattice constellations.

\textbf{Step 1b: Constellation partitioning}

A general method for constructing the overlapped constellation $\boldsymbol{ \mathcal S}$  is   to  partition a subset constellation from  a lattice constellation $\mathbf \Lambda$ that lies within some region $\mathcal R$. As such, the resultant  $\boldsymbol{ \mathcal S}$ is  a lattice code \cite{ForneyMultidimensional}. An important characteristic in a Lattice constellation is the constellation shape gain, which is defined as follows: 

\textit{Definition 3:  The shape gain  of a region $\mathcal R$ is defined as  \cite{ForneyMultidimensional}
\begin{equation}
\small
\gamma_{s}\ (\mathcal{R})=\frac{[V\ (\mathcal{R})]^{\frac{2}{n}}}{6  P(\mathcal R)}, 
\end{equation}
where $V(\mathcal{R})$ is the  volume of the region $\mathcal{R}$, $P(\mathcal R)$ denotes the average energy of the constellation and $n$ is the number of dimensions.}

%Note that the main source of excellent performance of SCMA is the high coding and  shape gain \cite{TaherzadehSCMA}.
%   A higher coding and shape gain indicates a more efficient use of the available signal space, resulting in better resistance to noise \cite{ForneyMultidimensional}.  These   observations  motivate us to further consider the diversity, coding and  shape gain of a codebook as the design metric of the proposed codebooks.

In general, a large shape gain is desirable for performance improvement \cite{ForneyMultidimensional}. Consequently, the shape gain is considered when in  determining the partition regions.  Overall, the  principles in determining the partition regions are given as follows:

\begin{itemize}
  \item  The partition region should be chosen  to enclose the desired number  of
signal points, i.e, $\vert \boldsymbol{ \mathcal S}\vert = M^{d_f}$.  
  \item A symmetric  partition region is preferable as the partitioned constellations are more likely symmetric and own a zero mean.
  \item  The partition region should be chosen to  maximize the shape gain of the constellation, i.e., $\gamma_{s}\ (\mathcal{R})$.
\end{itemize}

\begin{figure}
	\centering
	\begin{subfigure}{0.2 \textwidth}
  \includegraphics[width=1 \linewidth]{{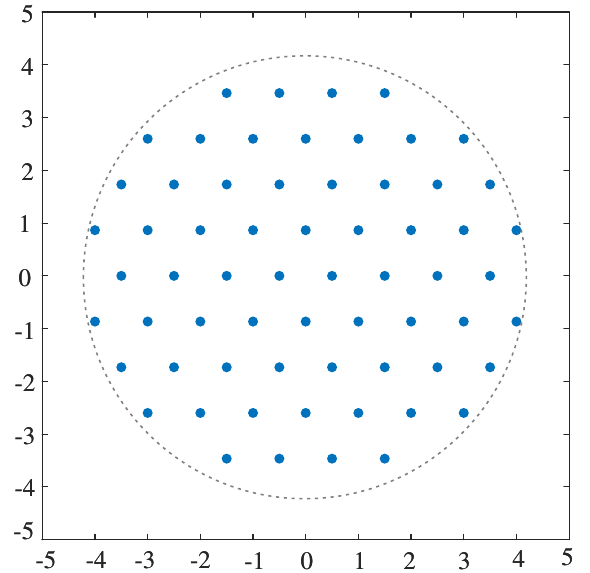}}
		\caption{ Eisenstein integers with circular window. }
	\end{subfigure}
	\begin{subfigure}{0.195\textwidth}
  \includegraphics[width=1 \linewidth]{{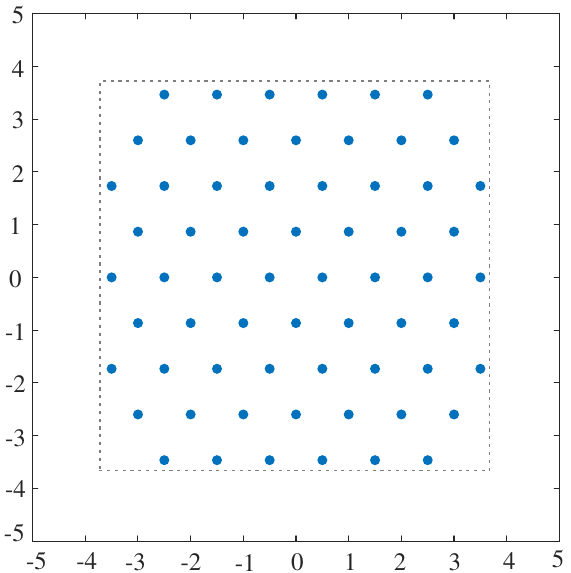}}
		\caption{Eisenstein integers with rectangular window.   }
	\end{subfigure}
	\begin{subfigure}{0.2 \textwidth}
  \includegraphics[width=1 \linewidth]{{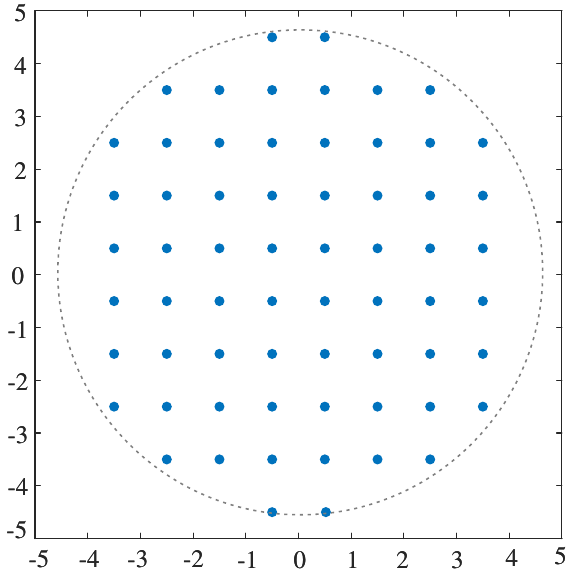}}
		\caption{  Gaussian integers with circular window. }
	\end{subfigure}
	\begin{subfigure}{0.2\textwidth}
  \includegraphics[width=1 \linewidth]{{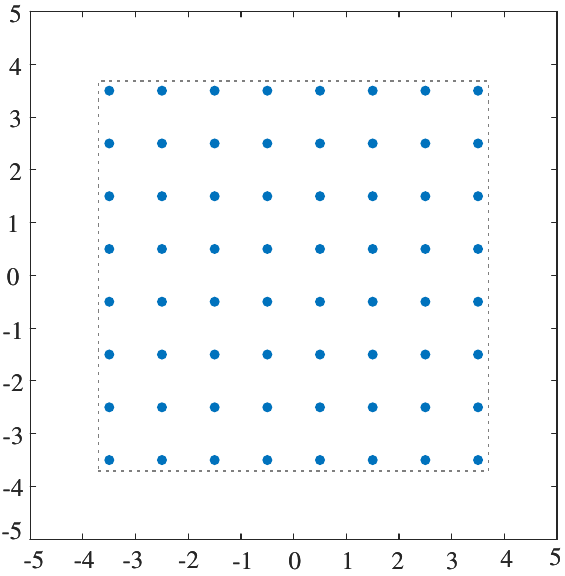}}
		\caption{  Gaussian integers with rectangular window.  }
	\end{subfigure}	
	\caption{Partitioned Lattice codes.    } 
	\label{Parti}
 \end{figure}

Considering the above partition rules, we employ two portion windows $\mathcal R$, namely the circular and rectangular  windows.  Fig. \ref{Parti} shows an example the  partitioned lattice codes based on the two windows for $M=4$ and $ d_f=3$. Compared to existing overlapped constellation, e.g., the overlapped constellations in Fig. \ref{LatticeCon}, the proposed constellation owns a good geometric shape and large MED.  

% The   peak energy of $\boldsymbol{ \mathcal S}$, denoted by $E_{\max}(\boldsymbol{ \mathcal S})$, is upper bounded  and approximated by the peak energy
%  of any point in the region $\mathcal R$. Hence, the PAPR can be  approximated by
% \begin{equation}
%     \small
%     \text{PAPR}(\boldsymbol{ \mathcal S}) \equiv \frac{E_{\max}(\boldsymbol{ \mathcal S})}{E_{\text{ave}}(\boldsymbol{ \mathcal S})} \approx \frac{E_{\max}({ \mathcal R})}{E_{\text{ave}}({ \mathcal R})}.
% \end{equation}

 \subsubsection{  Design the non-linear mapping} Upon determining  $\boldsymbol{ \mathcal S}$, the mapping between the incoming bits of $d_f$ users to the  $\boldsymbol{ \mathcal S}$ at   each subcarrier should be further designed.  Let 
 $\mathbf b = [ \mathbf b_1^{\mathcal T}, \mathbf b_2^{\mathcal T}, \ldots, \mathbf b_{d_f}^{\mathcal T}  ]^{\mathcal T}$ be the incoming message vector of the $d_f$  users with length of $L =d_f\log_2(M)$.  Then, the mapping process can be implemented with a permutation matrix,  which can be  mathematically expressed as  
   \begin{equation}
    \small
    \label{BitMpa}
    \overline {\mathbf b} = \mathbf P\mathbf b,
\end{equation}
where  $\mathbf P = \left\{ p_{l,l'} \right\} \in \mathbb B^{L \times L}  $ is the bit mapping (permutation) matrix and $p_{l,l'}=1$ denotes the $l'$th bit in $\mathbf b$ is mapped to the $l$th entry of $\overline {\mathbf b}$. Finally,  the constellation point with a labeling of $ \overline {\mathbf b}$, denoted by $s_{\overline {\mathbf b}} \in \boldsymbol{ \mathcal S}$,  is selected for transmission.  Moreover, since $\mathbf b$ contains $d_f$ users' bit message, we  further denote the mapping from the $i$th position of  $\mathbf b$, i.e., $\mathbf b_i$, to     $\overline {\mathbf b} $ in (\ref{BitMpa}) as
\begin{equation}
\small
    f_{\mathbf P}^{i}: \mathbf b \rightarrow \overline {\mathbf b},  1 \leq i\leq d_f.
\end{equation}
For example, for $M=4$ and
\begin{equation}
    \small
  \mathbf   P=\left[ \begin{matrix}
   0 & 0 & 1 & 0 &  0 & 0  \\
   1 & 0 & 0 & 0 & 0 & 0\\
   0 & 0 & 0 & 0 & 0 & 0 \\
   0 & 1 & 0 & 1 & 0 & 0  \\
   0 & 0 & 0 & 0 & 1 & 0  \\
   0 & 0 & 0 & 0 & 0 & 1  
\end{matrix} \right],
\end{equation}
$f_{\mathbf P}^{1}$ indicates that the incoming message bits, which are allocated to the first and second positions of $\mathbf b$, are mapped to the second and the fourth entries of $\overline {\mathbf b}$.   Considering the factor graph in (\ref{Factor_46}),    the following mapping rule is employed  to construct the NLCBs: 
  \begin{equation} 
 \label{mAPPING_46}
 \small
 {{f}}_{\mathbf F_{4 \times 6}}=\left[ \begin{matrix}
   0  & f_{\mathbf P}^{3} & f_{\mathbf P}^{1} & 0 &  f_{\mathbf P}^{2} & 0  \\
   f_{\mathbf P}^{2} & 0 &f_{\mathbf P}^{3} & 0 & 0 & f_{\mathbf P}^{1}\\
   0 &f_{\mathbf P}^{2}& 0 & f_{\mathbf P}^{1} & 0 & f_{\mathbf P}^{3} \\
   f_{\mathbf P}^{1} & 0 & 0 & f_{\mathbf P}^{2} & f_{\mathbf P}^{3} & 0  \\
\end{matrix} \right].
  \end{equation}
Based on (\ref{mAPPING_46}), the first user  employs the mapping principles $ f_{\mathbf P}^{2} $ and $ f_{\mathbf P}^{1} $ to map its input message to the second and fourth subcarrier, respectively, and the bits of $j$th user $j= 2, \ldots, J$  can be mapped in a similar way. For each RE, the mapping set is the same, leading to the same overlapped constellation $ \boldsymbol{ \mathcal S}$.

\begin{algorithm}[t] 
\caption{Generalized NLCB Design.}
\label{algorithm1}
\begin{algorithmic}[2]
\REQUIRE{  Nonlinear mapping $ {{f}}_{\mathbf F}$, indicator matrix $ {\mathbf F}$, $I_t$ and  partition window $\mathcal R$. } \\
\STATE   \textbf{Step 1} : Generate the lattice constellation.\\
\STATE   \textbf{Step 2} :  Apply a partition window to the lattice constellation to obtain the overlapped constellation  $ \boldsymbol{ \mathcal S}$. \\
 \STATE  \textbf{Step 3} : Maximizing $  \text{MED} \left ( \boldsymbol{\Phi}_{M^J} \right)$:\\
 \FOR{$i = 1:I_t$} 
\STATE     Randomly assign the bit mapping matrix and labeling. \\
\STATE     Preserve the current results of $  \text{MED}\left ( \boldsymbol{\Phi}_{M^J} \right)$.\\
 \ENDFOR \\
 \STATE Choose the best results of $\mathbf P$ and labeling that has the maximum value of $  \text{MED} \left ( \boldsymbol{\Phi}_{M^J} \right)$.\\
\end{algorithmic}
\end{algorithm}

 \subsubsection{MED optimization} 

In the proposed NL-SCMA, users' binary message  are directly mapped to a superimposed codeword for transmission. Obviously, the MED of the superimposed constellation, i.e.,  $\text{MED} \left ( \boldsymbol{\Phi}_{M^J} \right)$   depends on the ${{f}}_{\mathbf F_{4 \times 6}}$ and the bit labeling  of $ \boldsymbol{ \mathcal S}$.  In this regard, the mapping matrix $\mathbf P $  and the labeling rule of $ \boldsymbol{ \mathcal S}$ can be constructed to achieve a large $\text{MED} \left ( \boldsymbol{\Phi}_{M^J} \right)$. A generalized approach is to conduct random search to find      $\mathbf P $  and  labeling rule. \textbf{Algorithm 1} shows the proposed generalized MED optimization for the proposed NLCB.
It is worth emphasizing here that the codebook design is an offline process, and once the codebook is obtained, it can be directly applied to a practical system.  

In what follows we present a low complexity NLCB design which can avoid the random search. Specifically, an error-pattern inspired NLCB design is proposed to determine $\mathbf P $  and the labeling rule for $M=4$ and $d_f=3$. We begin with the analysis of the  $\text{MED} \left ( \boldsymbol{\Phi}_{M^J} \right)$  of the proposed NL-SCMA. 
   
  Let us rewritten   the instantaneous transmitted binary vector of $J$ users as a matrix based on the indicator matrix as follows 
  \begin{equation} 
 \label{TMb}
 \small
 \begin{aligned}
  {\mathbf{B}_{4 \times 6}} =\left[ \begin{matrix}
   \mathbf{0}  &  \mathbf{b}_{2} &  \mathbf{b}_{3}&\mathbf{0}   &   \mathbf{b}_{5} &   \mathbf{0} \\
    \mathbf{b}_{1} & \mathbf{0}  & \mathbf{b}_{3} &\mathbf{0}   & \mathbf{0}  &  \mathbf{b}_{6}\\
    \mathbf 0 & \mathbf{b}_{2}& \mathbf{0}  & \mathbf{b}_{4}&  0& \mathbf{b}_{6} \\
    \mathbf{b}_{1} & \mathbf{0}  & \mathbf{0}  & \mathbf{b}_{4}& \mathbf{b}_{5} & \mathbf{0}   \\
\end{matrix} \right].
 \end{aligned}
  \end{equation}
For an instantaneous transmitted $ {\mathbf{B}_{4 \times 6}}$, due to multiuser interference and additive white Gaussian noise, it is assumed to be
erroneously decoded to another binary  matrix at the receiver, denoted by
 $\widetilde{\mathbf{B}}_{4 \times 6}$, where $\widetilde{\mathbf{B}}_{4 \times 6} \neq {\mathbf{B}}_{4 \times 6} $.    The error patterns between $\widetilde{\mathbf{B}}_{4 \times 6} $ and $ {\mathbf{B}}_{4 \times 6} $ can be basically categorized into two patterns. In the case of the  decoding errors occur with a single user only, the error pattern is called the “single-user error pattern (SUEP)”;  otherwise, it will be called the “multiple-user error patterns (MUEPs)” \cite{liu2021sparse}.  The following shows   examples of a SUEP and     MUEPs, respectively,
  \begin{equation} 
 \label{TMb}
 \small
 \begin{aligned}
    {\widetilde{\mathbf{B}}_{4 \times 6}^{(\text{SUEP})}} =\left[ \begin{matrix}
   \mathbf{0}  &  \mathbf{b}_{2} &  \mathbf{b}_{3}&\mathbf{0}   &   \mathbf{b}_{5} &   \mathbf{0} \\
    \widetilde {\mathbf{b}}_{1} & \mathbf{0}  & \mathbf{b}_{3} &\mathbf{0}   & \mathbf{0}  &  \mathbf{b}_{6}\\
    \mathbf 0 & \mathbf{b}_{2}& \mathbf{0}  & \mathbf{b}_{4}&  \mathbf{0}& \mathbf{b}_{6} \\
    \widetilde {\mathbf{b}}_{1} & \mathbf{0}  & \mathbf{0}  & \mathbf{b}_{4}& \mathbf{b}_{5} & \mathbf{0}   \\
\end{matrix} \right],
 \end{aligned}
  \end{equation}
  and 
    \begin{equation} 
 \label{TMb2}
 \small
 \begin{aligned}
  {\widetilde{\mathbf{B}}_{4 \times 6}^{(\text{MUEPs})}} =\left[ \begin{matrix}
   \mathbf{0}  &  \mathbf{b}_{2} &  \mathbf{b}_{3}&\mathbf{0}   &   \widetilde {\mathbf{b}}_{5} &   \mathbf{0} \\
    \widetilde {\mathbf{b}}_{1} & \mathbf{0}  & \mathbf{b}_{3} &\mathbf{0}   & \mathbf{0}  &  \mathbf{b}_{6}\\
    \mathbf 0 & \mathbf{b}_{2}& \mathbf{0}  & \mathbf{b}_{4}&  0& \mathbf{b}_{6} \\
    \widetilde {\mathbf{b}}_{1} & \mathbf{0}  & \mathbf{0}  & \mathbf{b}_{4}& \widetilde {\mathbf{b}}_{5} & \mathbf{0}   \\
\end{matrix} \right].
 \end{aligned}
  \end{equation}
  It is noted that the error pattern in Gaussian channels directly relates to the MED features of the superimposed constellation. In high SNR region,  if the $\text{MED} \left ( \boldsymbol{\Phi}_{M^J} \right)$ is observed within a single user only, then the SUEP dominants the error rate performance of the codebook, otherwise, the MUEPs  dominates. To proceed, let us reformulate $\text{MED} \left ( \boldsymbol{\Phi}_{M^J} \right)$  as
\begin{equation} 
 \label{MEDCat}
 \small
 \begin{aligned}
 \text{MED} \left ( \boldsymbol{\Phi}_{M^J} \right) = \min \left\{ \text{MED}^{\text{SUEP}}\left ( \boldsymbol{\Phi}_{M^J} \right), \text{MED}^{\text{MUEPs}}\left ( \boldsymbol{\Phi}_{M^J} \right) 
 \right  \},
 \end{aligned}
  \end{equation}
where $\text{MED}^{\text{SUEP}}\left ( \boldsymbol{\Phi}_{M^J} \right)$ and $ \text{MED}^{\text{MUEPs}}\left ( \boldsymbol{\Phi}_{M^J} \right)$ denote the MED between the SUEP  and MUEPs, respectively. In the existing linear SCMA codebooks, e.g., \cite{Huang,chen2020design, yan2016top,XiaoCapacity,VikasSCMA,VikasComprehensive,Zhang},  the following generally holds 
\begin{equation}
\small
\label{MEDcom}
\begin{aligned}
 \text{MED}^{\text{SUEP}}\left ( \boldsymbol{\Phi}_{M^J} \right)  =  \underset { 1 \leq j \leq J}{\min} \text{MED} \left (\mathcal{X}_j \right)   > \text{MED}^{\text{MUEPs}}\left ( \boldsymbol{\Phi}_{M^J} \right),    
\end{aligned}
\end{equation}
which  indicates the    MUEPs  dominate  the  error rate performance. 
 For example, the codebooks in \cite{chen2020design} and \cite{huawei}  have the same $\text{MED}^{\text{SUEP}}\left ( \boldsymbol{\Phi}_{M^J} \right) =1.4$, whereas the $ \text{MED}^{\text{MUEPs}}\left ( \boldsymbol{\Phi}_{M^J} \right)$ are given by $1.07$ and $0.56$, respectively. Therefore, numeral efforts were paid to maximize the $\text{MED}^{\text{MUEPs}}\left ( \boldsymbol{\Phi}_{M^J} \right)$.  On the other hand, we show that the proposed NL-SCMA codebook owns a large lower bound of $ \text{MED}^{\text{MUEPs}}\left ( \boldsymbol{\Phi}_{M^J} \right) $.   Let us define the element-wise distance at the $k$th entry between the transmitted vector $\mathbf w$ and decoded vector $\widetilde {\mathbf w}$ as $\tau _{{{\mathbf w}} \rightarrow {\widetilde {\mathbf{w}}}}(k)$. For
for MUEPs in (\ref{TMb2}),  we have 
\begin{equation}
    \small
    \tau _{{{\mathbf w}} \rightarrow {\widetilde {\mathbf{w}}}}(k) \geq \text{MED} \left ( \boldsymbol{ \mathcal S} \right),   k \in \{1,2,4 \}.
\end{equation}

  Obviously,   $ \text{MED}^{\text{MUEPs}}\left ( \boldsymbol{\Phi}_{M^J} \right) $ is lower bounded by 
\begin{equation}
    \small
    \text{MED}^{\text{MUEPs}}\left ( \boldsymbol{\Phi}_{M^J} \right) \geq  (d_v+1) \text{MED} \left ( \boldsymbol{ \mathcal S} \right).
\end{equation}
Since the designed Lattice code owns  a large  $ \text{MED} \left ( \boldsymbol{ \mathcal S} \right)$, the $ \text{MED}^{\text{MUPs}}\left ( \boldsymbol{\Phi}_{M^J} \right)$ can be guaranteed to be a large value. Hence, the remaining  problem is   to find a labeling and mapping  rule that achieves a large  $ \text{MED}^{\text{SUEP}}\left ( \boldsymbol{\Phi}_{M^J} \right)$.

Recall (\ref{BitMpa}), for $M=4$ and $d_f =3$,    $\widetilde{\mathbf b}$ can be expressed as $\widetilde {\mathbf b} = [\widetilde {\mathbf b}_{\text{H}}^{\mathcal T}, \widetilde {\mathbf b}_{\text{M}}^{\mathcal T}, \widetilde {\mathbf b}_{\text{L}}^{\mathcal T}]^{\mathcal T}$, where $\widetilde {\mathbf b}_{\text{H}}\in \mathbb B^{\log_2(M)  \times 1}, \widetilde {\mathbf b}_{\text{M}} \in \mathbb B^{\log_2(M) \times 1} $ and $ \widetilde {\mathbf b}_{\text{L}} \in \mathbb B^{\log_2(M)  \times 1}$  denote the highest, medium and least significant bit layers, respectively.  Then, the MED of  $\widetilde {\mathbf b}_{\text{H}}$ can be expressed as 
   \begin{equation}
    \small
 {d}_{\min,\widetilde {\mathbf b}_{\text{H}}}^{\boldsymbol{ \mathcal S}} =   \underset{\substack{ \widetilde {\mathbf b}_{\text{H}}' \neq \widetilde {\mathbf b}_{\text{H}} \\
  \widetilde {\mathbf b}_{\text{M}}' = \widetilde {\mathbf b}_{\text{M}},
   \widetilde {\mathbf b}_{\text{L}}' = \widetilde {\mathbf b}_{\text{L}}
 }}{\min }   \vert s_{\widetilde {\mathbf b}} -  s_{ \widetilde{\mathbf b}'}  \vert^2.
\end{equation}
   Similarly, we can obtain the MED of  $\widetilde {\mathbf b}_{\text{M}}$ and $\widetilde {\mathbf b}_{\text{L}}$,  which are denoted as $ {d}_{\min,\widetilde {\mathbf b}_{\text{M}}}^{\boldsymbol{ \mathcal S}} $ and $ {d}_{\min,\widetilde {\mathbf b}_{\text{L}}}^{\boldsymbol{ \mathcal S}}$, respectively.

\begin{algorithm}[t] 
\caption{Error Pattern-Inspired NLCB Design.}
\label{algorithm2}
\begin{algorithmic}[2]
%\REQUIRE{  $\mathcal T_{M,T}$}\\
 \STATE  \textbf{Initialize}: $\mathbf P= \mathbf I_{L}$.\\
\STATE   \textbf{Step 1} : The constellations are divided equally into $M$ groups based on their quadrant. Then, each group is labeled with the same labeling at the HSBs.  \\
%$  {\gamma_{i}} \gets$ 
 \FOR{$m$th group} 
 \FOR{$i = 1:I_t$} 
\STATE     Label  the  HSBs and LSBs randomly.\\
\STATE     Preserve the current labeling if $  {d}_{\min,\widetilde {\mathbf b}_{\text{H}}}^{\boldsymbol{ \mathcal S}} \geq  {d}_{\min,\widetilde {\mathbf b}_{\text{M}}}^{\boldsymbol{ \mathcal S}} \geq \gamma_{\text{th}}$.\\
 \ENDFOR \\
 \ENDFOR \\
 \STATE  Construct the NLCB based on the  bit layer assignment matrix in (\ref{mAPPING_v2}). \\
\end{algorithmic}
\end{algorithm}

In the proposed error pattern-inspired codebook design, priority is assigned to the highest and medium significant bit layers. Our objective is to determine a labeling rule that maximizes  ${d}_{\min, \widetilde {\mathbf b}_{\text{H}}}^{\boldsymbol{ \mathcal S}} $ and $  {d}_{\min, \widetilde {\mathbf b}_{\text{M}}}^{\boldsymbol{ \mathcal S}}$. Specifically, the constellation is first  divided into  $M$ groups, denoted as   $\boldsymbol{ \mathcal S}(1),\boldsymbol{ \mathcal S}(2), \ldots, \boldsymbol{ \mathcal S}(M)$, respectively,   where each group consists of $M^{d_f -1}$ elements. In particular, it is preferable to assign the elements within the same quadrant to the same group.  Then, all of the signals in each group are labeled with the same bit values at the highest significant bit layer, as such,  we can obtain     more degree of freedoms for improving ${d}_{\min,\widetilde {\mathbf b}_{\text{M}}}^2$  due to the $M$ groups are well separated. Denote  $ {d}_{\min,\widetilde {\mathbf b}_{\text{M}}}^{\boldsymbol{ \mathcal S}(m)}$ by  the  MED of the medium significant bit layers in the $m$th group, which takes the following expression    
   \begin{equation}
    \small
 {d}_{\min,\widetilde {\mathbf b}_{\text{M}}}^{\boldsymbol{ \mathcal S}(m)} =   \underset{\substack{s_{\widetilde {\mathbf b}},s_{\widetilde {\mathbf b}'} \in \boldsymbol{ \mathcal S}(m),\\
 \widetilde {\mathbf b}_{\text{M}}' \neq \widetilde {\mathbf b}_{\text{M}}, 
   \widetilde {\mathbf b}_{\text{L}}' = \widetilde {\mathbf b}_{\text{L}}
 }}{\min }   \vert s_{\widetilde {\mathbf b}} -  s_{ \widetilde {\mathbf b}'}  \vert^2. 
\end{equation}
 Subsequently, the labeling of the medium significant bit layers at each layer  are carried out   by maximizing  $  {d}_{\min,\widetilde {\mathbf b}_{\text{M}}}^{\boldsymbol{ \mathcal S}(m)} $ while maintaining  ${d}_{\min,\widetilde {\mathbf b}_{\text{H}}}^{\boldsymbol{ \mathcal S}} $ larger than a threshold $\gamma_{\text{th}}$.  Then ${d}_{\min,\widetilde {\mathbf b}_{\text{M}}}^{\boldsymbol{ \mathcal S}}$ can be expressed as   
  \begin{equation}
  \small
   {d}_{\min,\widetilde {\mathbf b}_{\text{M}}}^{\boldsymbol{ \mathcal S}} =  \underset{1 \leq m \leq M}{\min} \;  {d}_{\min,\widetilde {\mathbf b}_{\text{M}}}^{\boldsymbol{ \mathcal S}(m)}.
\end{equation}
The detailed   steps of the proposed  error pattern-inspired NLCB design  are summarized in  \textbf{Algorithm \ref{algorithm2}}.   Since the priority is assigned    to $ {d}_{\min,\widetilde {\mathbf b}_{\text{H}}}^{\boldsymbol{ \mathcal S}} $ and $ {d}_{\min,\widetilde {\mathbf b}_{\text{M}}}^{\boldsymbol{ \mathcal S}}$, we have 
  \begin{equation}
    \small
  {d}_{\min,\widetilde {\mathbf b}_{\text{H}}}^{\boldsymbol{ \mathcal S}} \geq  {d}_{\min,\widetilde {\mathbf b}_{\text{M}}}^{\boldsymbol{ \mathcal S}} \geq {d}_{\min,\widetilde {\mathbf b}_{\text{L}}}^{\boldsymbol{ \mathcal S}} \geq   \text{MED} \left ( \boldsymbol{ \mathcal S} \right).
\end{equation}
To improve $\text{MED}^{\text{SUEP}}\left ( \boldsymbol{\Phi}_{M^J} \right)$ and achieve fairness between $J$ users, the following mapping rule is  considered:
  \begin{equation} 
 \label{mAPPING_v2}
 \small
 {{f}}_{\mathbf F_{4 \times 6}}=\left[ \begin{matrix}
   0  & f_{\mathbf P}^{1} & f_{\mathbf P}^{2} & 0 &  f_{\mathbf P}^{3} & 0  \\
   f_{\mathbf P}^{3} & 0 &f_{\mathbf P}^{2} & 0 & 0 & f_{\mathbf P}^{1}\\
   0 &f_{\mathbf P}^{3}& 0 & f_{\mathbf P}^{2} & 0 & f_{\mathbf P}^{1} \\
   f_{\mathbf P}^{1} & 0 & 0 & f_{\mathbf P}^{2} & f_{\mathbf P}^{3} & 0  \\
\end{matrix} \right],
  \end{equation}
where $\mathbf P =  \mathbf I_L $, $ f_{\mathbf P}^{1},  f_{\mathbf P}^{2}$, and $ f_{\mathbf P}^{3}$ denote the mapping  of  input message to the highest, medium and least significant bit layers, respectively. In (\ref{mAPPING_v2}), the user maps its binary bits to the least significant bit layers  of $\boldsymbol{ \mathcal S}$ at one subcarrier will assign its bits to  the highest significant bit layers  of $\boldsymbol{ \mathcal S}$ at another subcarrier. As such, $ \text{MED}^{\text{SUEP}}\left ( \boldsymbol{\Phi}_{M^J} \right) $   can be improved.     Based on (\ref{mAPPING_v2}), the $\text{MED}^{\text{SUEP}}\left ( \boldsymbol{\Phi}_{M^J} \right)$ is given by 
  \begin{equation}
    \small
   \text{MED}^{\text{SUEP}}\left ( \boldsymbol{\Phi}_{M^J} \right) = \min\left \{  {d}_{\min,\widetilde {\mathbf b}_{\text{H}}}^{\boldsymbol{ \mathcal S}}+  {d}_{\min,\widetilde {\mathbf b}_{\text{L}}}^{\boldsymbol{ \mathcal S}}, 2 {d}_{\min,\widetilde {\mathbf b}_{\text{M}}}^{\boldsymbol{ \mathcal S}}
   \right\}.
\end{equation}
Similarly,  the MPD of the proposed NLCB  is given by
  \begin{equation}
    \small
   \text{MPD} \left ( \boldsymbol{\Phi}_{M^J} \right) = \min\left \{  {d}_{\min,\widetilde {\mathbf b}_{\text{H}}}^{\boldsymbol{ \mathcal S}}  {d}_{\min,\widetilde {\mathbf b}_{\text{L}}}^{\boldsymbol{ \mathcal S}}, ( {d}_{\min,\widetilde {\mathbf b}_{\text{M}}}^{\boldsymbol{ \mathcal S}})^2
   \right\}.
\end{equation}

To sum up, we have proposed a generalized architecture of designing the NLCBs based on lattice codes. Then an error pattern-inspired NLCB design is further proposed to simply the bit labeling and mapping design.

  \begin{table*}[htbp]
  \small
 \centering
 \caption{Comparison of the KPIs of  different codebooks. ``Gau.'' and ``Ray.'' denote the Gaussian and Rayleigh fading channels, respectively.}
     \begin{tabular}{c|c|c|c|c|c}
    \hline
    \hline
        {Codebook}    &  Target Channels  &  $ \text{MED} \left ( \boldsymbol{\Phi}_{M^{{J}}} \right)$ &  $ \text{MED} \left ( \boldsymbol{\Phi}_{M^{{d_f}}}(k) \right)$  & MPD  & \makecell{Shape gain, $\gamma_{s}\ (\mathcal{R})$}  \\
    \hline
    \hline
   Prop.  NLCB1 & Gau.\& Ray.   & 0.94  & 0.413& 0.61 & 5.90\\
    \cline{1-6}
  Prop.  NLCB2   & Gau.\& Ray.  & 0.92 & 0.410 &0.60 & 4.65\\
   \cline{1-6}
  Prop. NLCB3  & Gau.\& Ray.    & 1.02 & 0.383& 0.58  & 6.21 \\
      \cline{1-6}
  Prop.   NLCB4  & Gau.\& Ray.   & 1.07& 0.378 &0.58  &5.90 \\
      \cline{1-6}
  Huawei \cite{huawei}  & Gau.\& Ray.  & 0.56 & 0.146 &0.85  &- \\
      \cline{1-6}
  GAM  \cite{mheich2018design}   & Gau.\& Ray.  & 0.56 &0.000& 0.69  &- \\
      \cline{1-6}
  Chen-G  & Gau.   & 1.07 & 0.016 & 0.78 &-  \\
      \cline{1-6}
  Chen-R  & Ray.   & 0.84 & 0.199  & 1.0 & - \\
   %   \cline{1-7}
  %Star-QAM \cite{yu2015optimized}  & Gau. \& Ray. & 5.8 dB  & 0.90 &0.083 &0.72& 0  \\
    \hline
    \hline
     \end{tabular}
     \label{codebooksKPI}
 \end{table*}

 \section{Numerical Results}

In this section, we conduct numerical simulations to evaluate the proposed NLCBs.  The proposed NLCBs   based on Lattice codes shown in Figs. \ref{Parti}(a)-(d) are named as   NLCB1, NLCB2, NLCB3, NLCB4, respectively.    For comparison, we consider  the   GAM   codebook \cite{mheich2018design},   Huawei codebook  \cite{huawei},   and  Chen's codebook \cite{chen2020design}\footnote{It was shown in \cite{Zhang} that the Huawei codebook and GAM codebook outperform  many  state-of-the-art codebooks in terms of the coded  BER  performance. Additionally, the Chen codebook exhibits   promising uncoded BER performance  \cite{chen2020design}. Consequently, for   demonstrate the superiority of the proposed codebooks, we  consider the Huawei codebook, GAM codebook, and Chen codebook as the primary benchmark codebooks for the BER comparison.  }.  Specifically, the codebooks  designed for Gaussian and Rayleigh fading channels in \cite{chen2020design}  are  named as Chen-G and Chen-R, respectively.   Moreover, we consider a Rician fading channels, i.e.,  ${{h}_{k}} \sim \mathcal{CN}\left ({\sqrt{ \frac {\kappa}{1+\kappa}}, \sqrt{ \frac {1}{1+\kappa}}  }\right )$ with  $\kappa$ represents the ratio of average power in the  LoS path  over that in the scattered component. For $\kappa = 0$ and $\kappa \rightarrow \infty$, the Rician fading channels are the Gaussian and Rayleigh fading channels, respectively.
% StarQAM codebook \cite{yu2015optimized} 

\subsection{KPIs comparison}

Table \ref{codebooksKPI}  compares    the KPIs of  various codebooks.  One can see that the     proposed codebooks own large     $ \text{MED} \left ( \boldsymbol{\Phi}_{M^{{d_f}}}(k) \right) $   than   other codebooks. In particular, the proposed NLCB1 achieves the largest   $ \text{MED} \left ( \boldsymbol{\Phi}_{M^{{d_f}}}(k) \right) $. Moreover, the proposed NLCBs with circular window own  slightly larger  value of    $ \text{MED} \left ( \boldsymbol{\Phi}_{M^{{d_f}}}(k) \right) $ than that of the codebook obtained with a  rectangular window.  In addition, the proposed  NLCB4 codebooks  own  the same  MED with the Chen-G codebook. However, the proposed   NLCB4 codebooks can attain better KPI performances in terms of   $ \text{MED} \left ( \boldsymbol{\Phi}_{M^{{d_f}}}(k) \right) $, MPD and diversity than the Chen-G codebook.

\subsection{Uncoded BER performance}

  \begin{figure}[htbp]
     \centering
  \includegraphics[width=0.52 \textwidth]{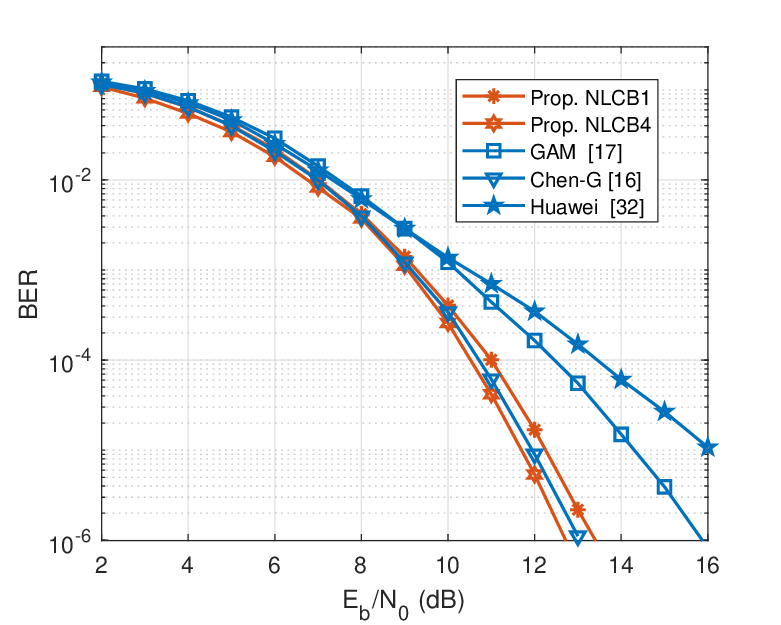}
         \caption{BER performance of various codebooks over Gaussian channels.}
         \label{FigSim_awgn}
\end{figure}
\begin{figure}[htbp]
     \centering
\includegraphics[width=0.52\textwidth]{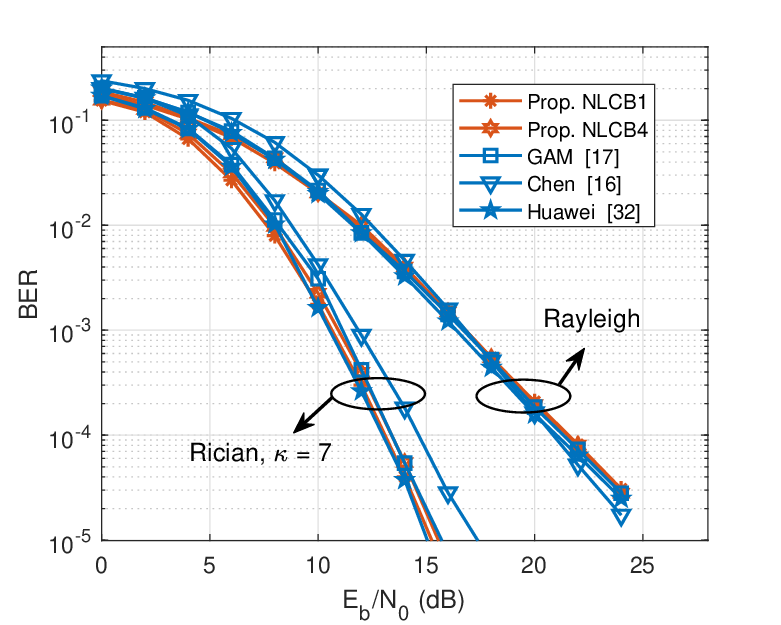}
         \caption{  BER performance of various codebooks over Rician fading channels ($\kappa = 0 $ and $\kappa = 7$).}
         \label{FigSim_ray}
\end{figure}

\begin{figure}[htbp]
     \centering
\includegraphics[width=0.5\textwidth]{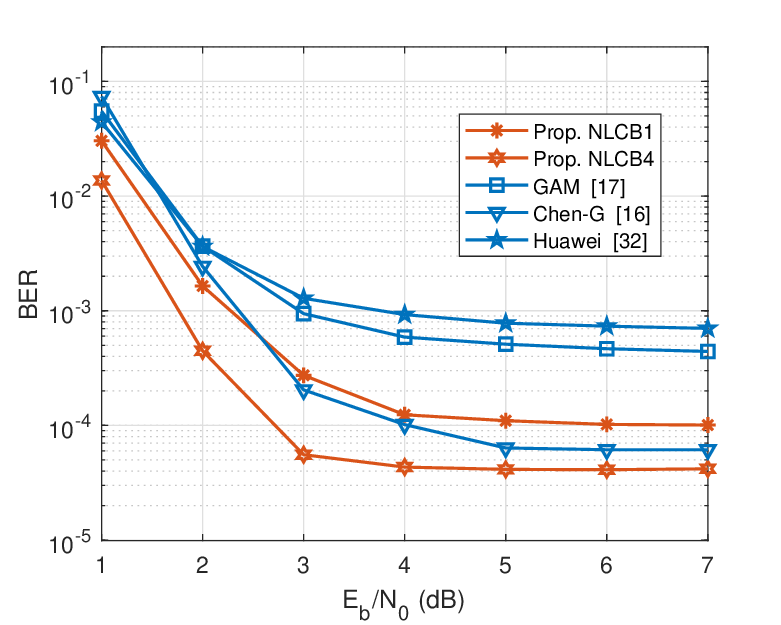}
         \caption{ BER performance v.s. number of MPA iterations}
         \label{GfigMPA}
\end{figure}

Fig. \ref{FigSim_awgn} and Fig. \ref{FigSim_ray} compare the   BER performance of various codebooks in an uncoded system. Specifically, the proposed  NLCB1 and NLCB4 are employed,  where the MPA iterations is $I=7$. The Chen-G is considered for Gaussian channels, whereas the Chen-R is employed  in Fig. \ref{FigSim_ray}.
As can be seen from the figures, the proposed codebooks achieve promising BER performance in   Gaussian, Rayleigh and Rician ($\kappa = 7$) fading channels. In Gaussian channels, the proposed NLCB1 achieves about $4$ dB gain over the Huawei codebook at the BER of $10^{-5}$. Although the MED of the proposed NLCB1  is the same with Chen codebook, the proposed NLCB1  still outperform the Chen codebook slightly. The main reason is that for the proposed codebook, we have $\text{MED} \left ( \boldsymbol{\Phi}_{M^J} \right) =   \text{MED}^{\text{SUEP}}\left ( \boldsymbol{\Phi}_{M^J} \right) $, whereas $\text{MED} \left ( \boldsymbol{\Phi}_{M^J} \right) =   \text{MED}^{\text{MU}} \left ( \boldsymbol{\Phi}_{M^J} \right)$ holds for the Chen-G codebook. Namely, the SUEP and MUEPs dominant  the BER performance of the proposed codebooks and the Chen-G codebook, respectively. It is evident that  the SUEP has smaller Hamming distance  than that of MUEPs. Hence, the proposed   NLCB1     can  attain better BER performance. 

Observed form Fig. \ref{FigSim_ray}, all the codebooks achieve a similar BER performance at the medium  SNRs over Rayleigh fading channels.      It is worth  mentioning that the proposed codebooks achieve better performance than the benchmark  codebooks in the low SNR regions, whereas the Chen-R codebook performs the best at the high SNR regions in Rayleigh fading channels. This is reasonable, as lower bounding the MED of all superimposed signal vectors helps improve performance in the low-SNR region, while improving the MPD is beneficial for enhancing error rate performance in the high-SNR region for downlink Rayleigh fading channels. The proposed codebooks achieve a similar BER performance with GAM and Huawei codebook in the medium-to-high SNR regions in the Rician fading channels of $\kappa = 7$. Similar to their performance in Rayleigh fading channels, the proposed codebooks outperform other codebooks in low SNR regions.

 Fig. \ref{GfigMPA} shows the BER perforamce  with different number of MPA iterations of various codebooks in Gaussian channels.   
As can be seen from the figure, similar to the beach-marking codebooks, the proposed codebooks also require  about $4$   MPA iterations to converge.

 \begin{table}  
\small
    \caption{Simulation Parameters}
    \centering
    \begin{tabular}{c|c}
    \hline
     \hline
       \textbf{ Parameters}  &  \textbf{ Values}  \\
        \hline
         \hline
       Transmission  &  Downlink \\
        \hline
       Channel model  &  Gaussian and Rayleigh fading channels \\
        \hline
        SCMA setting & $\lambda = 150\%$, $K=4,J=6$ \\
          \hline
        Channel coding &  \makecell[c] {  5G NR LDPC codes with code length  $=512$, \\ and     rate $=0.75$ (Fig. \ref{LDPC1}) and $0.50$ (Fig. \ref{LDPC2}). }  \\ 
         \hline
        Codebooks &  \makecell[c] {NLCB1, NLCB4, \\ GAM, Chen-G, Chen-R, and Huawei codebooks}  \\ 
       \hline
       Receiver &   \makecell[c] { Turbo-MPA: \\ $3$ MPA  iterations, $10$ LDPC iterations,  \\ and  4 outer iterations.  } \\
         \hline
    \end{tabular}
    \label{sim_para}
 \end{table}

\subsection{Coded BER performance}

In this subsection, we evaluate the  performance of different codebooks in a bit-interleaved coded modulation (BICM) system. Specifically, the iterative Turbo receiver in \cite{liu2021sparse} is employed.  At the receiver side,   Turbo decoding is
carried out between MPA decoder and channel decoder
 by iteratively exchanging soft information
in the form of log-likelihood ratio (LLR) including \textit{a priori}
LLR and \textit{a posterior} LLR (extrinsic).

\begin{figure}
     \centering
  \includegraphics[width=0.52 \textwidth]{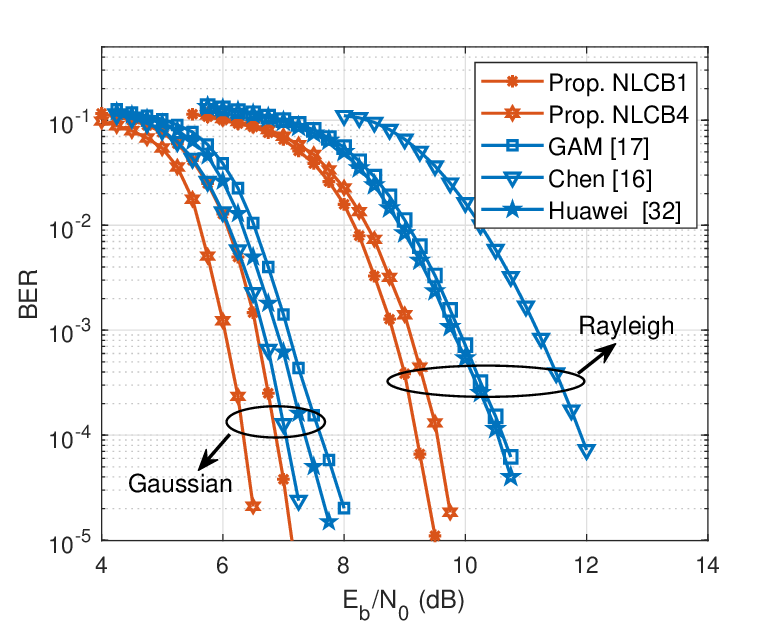}
              \caption{ LDPC coded BER performance of various codebooks with rate of $0.75$.}
              \label{LDPC1}
\end{figure}

\begin{figure}
     \centering
  \includegraphics[width=0.52 \textwidth]{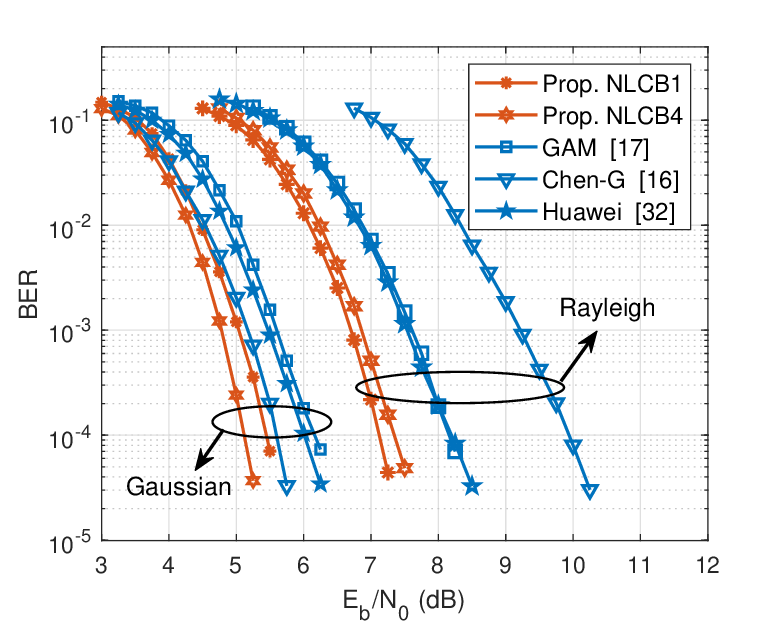}
              \caption{ LDPC coded BER performance of various codebooks with rate of $0.5$.}
              \label{LDPC2}
\end{figure}

\begin{figure}
     \centering
  \includegraphics[width=0.52 \textwidth]{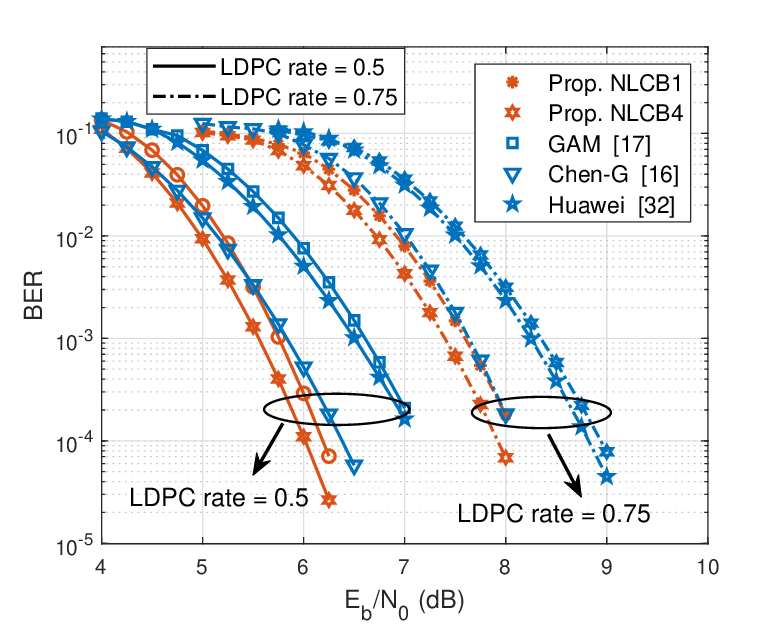}
              \caption{ LDPC coded BER performance of various codebooks with  $\kappa = 7$.}
              \label{LDPC3}
\end{figure}

 Figs. \ref{LDPC1}-\ref{LDPC3} compare  the  BER performance of the LDPC coded BICM systems. The detailed simulation parameters are summarized in Table \ref{sim_para}.  As can be seen from Fig. \ref{LDPC1}, the proposed   NLCB4 achieves about $1$ dB gain over the Chen-G codebook in Gaussian channels at BER of $10^{-4}$, whereas the proposed NLCB1 outperforms the Huawei and GAM codebook by   $1.5$ dB over   Rayleigh fading channels. It should be noted that the proposed codebooks can performance well in both Gaussian and Rayleigh fading channels, however, this may not hold for the benchmark codebooks.     Another observation is that  a codebook that achieves better BER performance in an uncoded system may not outperform in a BICM  system, and vice versa, which is also pointed out in our previous work  \cite{SSDSCMA}.

We further compare the coded BER performance with rate of $0.5$, which is shown in  Fig. \ref{LDPC2}. The gains between the proposed codebooks and the benchmark codebooks are smaller compared to that of  rate of $0.75$ in Fig. \ref{LDPC1}. However,   the proposed NLCB1 can still outperform  the Huawei and GAM codebook by   $1$ dB over   Rayleigh fading channels at the BER of $10^{-4}$. In addition,  the proposed     NLCB4 can   achieve about $0.5$ dB gain over the Chen codebook in Gaussian channels at the BER of $10^{-4}$.   In addition, Fig. \ref{LDPC3} shows the coded BER performance of different codebooks for Rician fading channels with $\kappa = 7$. The proposed NLCB1 still exhibits  the best BER performance in Rician fading channels.  It is also interesting to observe that, although the Chen-G codebook is designed for Gaussian channels, it significantly outperforms the GAM and Huawei codebooks in Rician fading channels.   Overall,  the benefits  of the proposed  codebooks are more prominent in a coded system.

\section{Conclusion  }

 In this paper, we have proposed a novel class of  NL-SCMA codebooks. Different from the existing SCMA schemes, where the incoming bits are first mapped to a sparse codeword and then superimposed for transmission,  the proposed NL-SCMA  directly maps  the incoming bits to a superimposed codeword.
 Despite the widely employed design KPIs in previous SCMA works, such as MED and MPD, we have also introduced the diversity and shape gain of a codebook as the design KPIs.  For efficient codebook design, different from the existing  ``top-down'' based design principles, we have  proposed a  ``down-top'' based  scheme to construct the   NLCBs. Specifically, we have adopted the Lattice codes as the superimposed constellation by partitioning a Lattice constellation. As such, large shape gain can be   guaranteed due to the inherent superiority of the Lattice constellations.  Upon determining the one dimensional Lattice code, a generalized NLCB design     has been  proposed to design the mapping between the input message and superimposed constellation by improving the  $\text{MED} \left ( \boldsymbol{\Phi}_{M^J} \right) $. In addition, by carefully analyzing the error patterns in the NL-SCMA systems, an error pattern-inspired NLCB design has been proposed to improve the  $\text{MED} \left ( \boldsymbol{\Phi}_{M^J} \right) $ while can significantly reduce the computational complexity during optimization.   Numerical results shown that the proposed NLCBs can  simultaneously achieve promising BER performance over both Gaussian and Rayleigh fading channels in  uncoded and coded systems, whistling achieving  full diversity and large shape gain. In particular, significant BER gain has been observed of the proposed codebooks for the BICM system.

\appendices

   \section{The proposed codebooks} 
   
\begin{figure}[htbp]
     \centering
  \includegraphics[width=0.42 \textwidth]{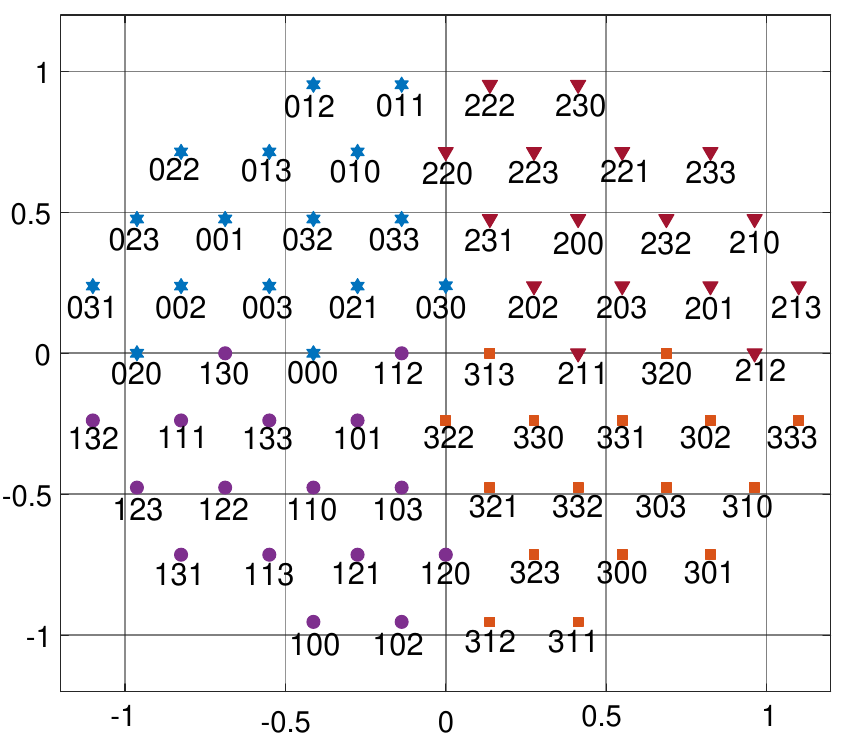}
         \caption{ The overlapped constellation $\boldsymbol{ \mathcal S}$ of  NLCB1.}
         \label{FigNLCB1}
\end{figure}

The proposed  $\boldsymbol{ \mathcal S}$ of  NLCB1 is shown in Fig. \ref{FigNLCB1}. For the simplify of presentation, the bit labeling of ``00", ``01", ``10", ``11"  in each bit layer of $\widetilde {\mathbf b}  $ are represented by the decimal numbers ``0", ``1", ``2", ``3", respectively. For example, the constellation point labelled with $\widetilde {\mathbf b} = [ 0, 0, 0, 1, 1, 0],$ i.e., $ \widetilde {\mathbf b}_{\text{H}} =[0, 0], \widetilde {\mathbf b}_{\text{M}}= [0, 1], \widetilde {\mathbf b}_{\text{L}} =[ 1, 0]$, is represented by ``012''.
  NLCB1 and NLCB4 can be constructed based on  Fig. \ref{FigNLCB1} with  the mapping defined in (\ref{mAPPING_v2}) and $ \mathbf W = \text{diag} (\mathbf I_J)  $.

  \label{cb}
\ifCLASSOPTIONcaptionsoff
  \newpage
\fi



%\bibliographystyle{IEEEtran}

\bibliography{ref} % put your favourite Bibtex archive references here
\bibliographystyle{IEEEtran}

% needed around the contents of the optional argument to biography to prpattern

% \end{IEEEbiography}


% % insert where needed to balance the two columns on the last page with
% %\newpage



%\vfill


\end{document}